\documentclass[11pt]{article}

\usepackage[a4paper,margin=1in]{geometry}
\usepackage{jheppub}
\usepackage{amsmath,amssymb,mathtools,physics,bm}
\usepackage{tensor}
\usepackage{slashed}
\usepackage{color}
\usepackage{comment}
\usepackage{natbib}
\usepackage{graphicx}
\usepackage{tikz}
\usetikzlibrary{arrows.meta,decorations.markings}

\newcommand{\hk}{\hat{k}}

\newcommand{\he}{\hat{\varepsilon}}

\numberwithin{equation}{section}
\usepackage[colorlinks=true,
            linkcolor=blue,
            citecolor=blue,
            urlcolor=blue]{hyperref}
\begin{document}

\title{Towards a recursive construction of string scattering amplitudes}
\author{Arkajyoti Manna}
\affiliation{Center for Fundamental Physics, School of Physical Science and Technology,\\
ShanghaiTech University, 393 Middle Huaxia Road, Shanghai 201210, China}
\emailAdd{amanna@shanghaitech.edu.cn} 

\abstract{We construct open superstring scattering amplitudes from three-point data using the Britto–Cachazo–Feng–Witten (BCFW) recursion and four-dimensional spinor helicity variables. We compute the contributions of selected leading and subleading Regge families to the color-ordered four-gluon amplitude by sewing their cubic couplings and summing over intermediate polarizations. The four-dimensional description makes helicity selection rules explicit and reduces the massive polarization sums to SU(2) little group contractions. The resulting residues are expressed in terms of Jacobi and Legendre polynomials, yielding explicit sums over mass levels. We construct a differential operator relating the two polynomial kernels, including their level-dependent normalizations. These results provide a concrete starting point for organizing the infinite intermediate-state sum and investigating whether relations among Regge families can support a recursive reconstruction of the full string amplitude.}

\keywords{String amplitudes, BCFW recursion, Regge trajectories.}

\maketitle

\section{Introduction}

The worldsheet formulation provides a systematic prescription for string scattering amplitudes \cite{Green:1987sp,Polchinski:1998rq,Friedan:1985ge,DHoker:1988pdl}, but their explicit evaluation requires integrating correlators of vertex operators over the insertion points. These integrals become increasingly complicated as the number of external particles grows \cite{Medina:2002ge,Oprisa:2005wu}. Even when the kinematic dependence is organized into compact expressions, the remaining multidimensional integrals present a substantial computational challenge \cite{Mafra:2011nv,Mafra:2011nw}. 
This motivates a complementary construction based on on-shell recursion \cite{Britto:2004ap,Britto:2005fq}, in which higher-point amplitudes are assembled from lower-point amplitudes through factorization. In string theory, the central challenge then becomes organizing and evaluating the sum over the infinite tower of intermediate states \cite{Boels:2008fc,Cheung:2010vn,Boels:2010bv,Fotopoulos:2010cm,Boels:2012if,Chang:2012qs}.

The success of on-shell recursion in gauge theory \cite{Britto:2004ap,Britto:2005fq} and gravity \cite{Bedford:2005tw,Cachazo:2005ca,Benincasa:2007qj,Arkani-Hamed:2008bsc} motivates its application to string scattering. Previous work has established the applicability of BCFW recursion to tree-level string amplitudes \cite{Boels:2008fc,Cheung:2010vn,Boels:2010bv} and explored several implementations \cite{Fotopoulos:2010cm,Fotopoulos:2010ay,Chang:2012qs,Srisangyingcharoen:2024qyx}. A central practical challenge is the infinite spectrum of intermediate states: recursion requires both a sum over mass levels and a sum over the physical states at each level \cite{Fotopoulos:2010cm,Chang:2012qs}. Turning this formal construction into an efficient calculational method therefore requires a manageable description of the cubic couplings and their polarization sums \cite{Sagnotti:2010at,Schlotterer:2010kk,Boels:2012if,Arkani-Hamed:2017jhn}.

String field theory provides a complementary framework for constructing string amplitudes, beginning with Witten’s cubic open bosonic theory \cite{Witten:1985cc} and its derivation of the Veneziano amplitude by Giddings \cite{Giddings:1986iy}. Subsequent developments constructed
consistent superstring interactions \cite{Erler:2013xta} and established
agreement with the perturbative tree-level S-matrix \cite{Konopka:2015tta}.
More recently, a construction inspired by this framework 
combines unintegrated vertex operators through a point-split product satisfying weak associativity, yielding recursive series representations without explicit worldsheet moduli integration \cite{Gomez:2026exq}. These developments provide complementary approaches to organizing the contributions of the infinite string spectrum.

In this work, we investigate this problem through selected leading Regge (LR) and subleading Regge (SLR) contributions to the color-ordered four-gluon amplitude. Starting from three-point amplitudes obtained using worldsheet vertex operators, we restrict the external kinematics to four dimensions and express the relevant massive polarizations in spinor helicity variables. This makes the four-dimensional spin content explicit and allows the intermediate polarization sums to be evaluated through SU(2) little-group contractions. The helicity decomposition also exposes selection rules in the cubic couplings, simplifying the factorization channels that contribute. The boundary analysis applies to the full string amplitude, while our LR and SLR calculations select particular families from its intermediate-state sum.

The resulting contributions are organized into kernels involving Jacobi and Legendre polynomials. We derive a differential relation between these kernels after separating their propagators and accounting for their normalization and kinematic arguments. This relation suggests a possible way to organize contributions from additional Regge families. The four-point calculation thus serves as a controlled setting in which to develop a recursive construction from three-point data, with the longer-term aim of reconstructing the full string amplitude and extending the method to higher multiplicity and massive external states. Relations proposed for the remaining families can be tested against the known four-point worldsheet amplitude.

This paper is organized as follows. In section \ref{sec:Review: validity of recursion relations}, we review the validity of the BCFW recursion relations in the context of open superstring scattering amplitudes. In section \ref{sec:Three-point amplitudes at subleading Regge}, we construct the required cubic coupling of two gluons and a massive state on a selected SLR trajectory. Section \ref{sec:Four gluon residue at leading Regge} derives the four-dimensional reduction of the known cubic coupling with a massive LR state, computes the four-gluon residue via BCFW recursion, and analyzes its high-energy behaviour using a novel contour representation. In section \ref{sec:Four gluon residue at subleading Regge}, we dimensionally reduce the ten-dimensional amplitude from section \ref{sec:Three-point amplitudes at subleading Regge} to four dimensions, construct the corresponding four-point residue via BCFW recursion, and examine its Regge behaviour using an alternative method. We conclude in section \ref{sec:conclusion} by establishing the relation between leading and subleading Regge kernels. The RNS superstring OPEs are summarized in appendix \ref{app:OPE_RNS_superstrings}. The remaining appendices cover the Pomeron vertex derivation (appendix \ref{app:Pomeron_vertex}), vanishing of the same-helicity sector in the four-dimensional LR cubic amplitude (appendix \ref{app:same_helicity_LR}), derivation of general rank-$n$ $\mathrm{SU}(2)$ contraction formula (appendix \ref{app:SLR_gluing}), technical details of the LR Regge limit (appendix \ref{app:Regge limit of LR contribution}), and the algebra relating the two Regge kernels (appendix \ref{app:kernel_relations}).

\section{ Review: validity of recursion relations}\label{sec:Review: validity of recursion relations}

The applicability of on-shell recursion techniques requires control over the amplitude at large complex deformation parameter $z$. In Yang-Mills theory, Arkani-Hamed and Kaplan interpreted this limit as the propagation of hard particles (with complex momentum $\hat{k}(z)$) through a soft background \cite{Arkani-Hamed:2008bsc} consisted of undeformed particles. The leading interaction in $z$ in this regime exhibits an enhanced \textit{spin-Lorentz} symmetry acting on the tensor indices of hard particles, while subleading interactions break this symmetry in a specific manner. Together with Ward identities, the resulting tensor structure determines the large-$z$ behaviour of the amplitude.

In string theory, the corresponding structure can be extracted directly from the worldsheet operator product expansion of vertex operators of the two deformed external states. Their integrated OPE is organized by the \textit{Pomeron vertex operator}, which isolates the leading dependence on the large complex momentum. The tensor structures accompanying this operator reveal the string counterpart of the enhanced spin-Lorentz symmetry and its breaking at subleading order in $z$ at large $z$. Here, we review this construction, introduced by Brower, Polchinski, Strassler and Tan \cite{Brower:2006ea} and applied to string BCFW recursion by Cheung, O'Connell and Wecht \cite{Cheung:2010vn}.

 We begin with the color-ordered $N-$gluon amplitude in open superstring theory in the standard worldsheet formulation \cite{Schlotterer:2011psa}\footnote{Here we suppress the helicity $h_i$ labels.}
	\begin{align}
		\hat{\mathcal{A}}_N(z) =  \prod_{i=2}^{N-2} \int_{\mathcal{D}} \mathrm{d} w_i \langle cV^{(-1)}_{1}(\hat{k}_1;w_1)\, V^{(-1)}_{2}(\hat{k}_2;w_2)\, V^{(0)}_3(k_3;w_3)\cr 
        \cdots cV^{(0)}_{N-1}(k_{N-1};w_{N-1}) cV^{(0)}_N(k_N;w_N)\rangle \,,
	\end{align}
where $z$ is the BCFW parameter defined below \eqref{BCFW_mom_shift} and the domain of worldsheet integration is
\begin{align}
    \mathcal{D}:\quad 0< w_2<\cdots w_{N-2}< 1\,. 
\end{align}
Here we have fixed the SL$(2,\mathbb{R})$ invariance by fixing the locations of punctures labeled by $i=1,N-1,N$ at $w_1=0,w_{N-1}=1,w_N=\infty$.  $V^{(p_i)}_i(k_i,w_i)$ denotes the vertex operator including both matter and superghost sectors with reparameterization ghosts $c(w)$ written separately, representing the $i$-th particle with momentum $k_i$ inserted at $w_i$ on the string worldsheet in the picture number $(p_i)$. Cancellation with the background superghost charge in the RNS formulation then impose $\sum_{i=1}^N p_i = -2$.
Moreover, $V_j(\hat{k}_j,w_j)$ for $j=1,2$, denotes the vertex operators with complex momentum to apply the on-shell recursion.
We take the complex momenta as\footnote{We use $(A,B,\cdots)$ to denote the 10d indices and ($\mu,\nu,\cdots$) in the 4d discussion later.}
 \begin{align}\label{BCFW_mom_shift}
\hat{k}^A_1=k^A_1 - z r^A \,, \qquad \hat{k}^A_2=k^A_2 + z r^A  \,,\qquad  r^2=0\,, 
 \end{align}
with $\hat{k}_{j}^2=k_{j}^2=k_{j}\cdot r=0 $. We work in the Minkowski signature and thus the BCFW parameter $z$ is complex.

To analyze the large $z$ behaviour of this $N-$point amplitude, we define the following \textit{integrated vertex pair}
\begin{align}\label{eq:two_point_amp}
    \widehat{\mathcal{V}}_{12}(\hat{\varepsilon}_1,\hat{\varepsilon}_2;z,w_3) := c(0)\, \int_0^{w_3} dw  V^{(-1)}_{1}(\hat{\varepsilon}_1,\hat{k}_1;0)\, V^{(-1)}_{2}(\hat{\varepsilon}_2,\hat{k}_2;w)\,.
\end{align}
We made the dependence on the deformed massless polarization explicit here.
Its large-$z$ behaviour is described by the local Pomeron insertion derived below.
With this, we rewrite $\hat{\mathcal{A}}_n(z)$ as 
\begin{align}\label{eq:N_point_type1_amp}
    \hat{\mathcal{A}}_N(z) = \prod_{i=3}^{N-2} \int_{\mathcal{D}'}  d w_i  \langle  \widehat{\mathcal{V}}_{12}(\hat{\varepsilon}_1,\hat{\varepsilon}_2;z,w_3)\,  V^{(0)}_3(k_3;w_3) \cdots cV^{(0)}_{N-1}(k_{N-1};1)\, cV^{(0)}_N(k_N;\infty)\rangle\,,
\end{align}
where the domain of worldsheet integration is
\begin{align}
    \mathcal{D}': \quad 0< w_3<\cdots w_{N-2}< 1\,. 
\end{align}
The entire dependence of the BCFW parameter $z$ is now carried in the operator $\widehat{\mathcal{V}}_{12}$.

To prove that BCFW recursion relation can be applied to construct the gluon amplitude, we now consider the vertex operator, in the above equation for massless gauge bosons in type-I superstring theory. The vertex operator in the canonical $(-1)$ picture is given by
\begin{align}\label{eq:gluon_vertex_op_can_pic}
	V_i^{(-1)}(\varepsilon_i,k_i;z_i)=\varepsilon_{i,A} \psi^A(z_i) e^{i k_i \cdot X(z_i)}e^{-\phi(z_i)} \,,
\end{align}
where we have stripped off the Chan-Paton factor.
We use the OPEs between various matter fields in appendix \ref{app:OPE_RNS_superstrings} to derive the OPE between the two deformed massless gauge boson vertex operators, keeping subleading terms in $w$ as follows
\begin{align}\label{eq:pomeron_VV_ope}
     V_1^{(-1)}(\hat{\varepsilon}_1,\hat{k}_1;0)\, & V_2^{(-1)}(\hat{\varepsilon}_1,\hat{k}_2;w) =  w^{\alpha' s_{12}-2} e^{ik \cdot X+iw \hat{k}_2\cdot \dot{X}} e^{-2\phi} \cr
     & \times \Big[  (\hat{\varepsilon}_1 \cdot \hat{\varepsilon}_2) \left(1 -w \dot{\phi} + \frac{i w^2}{2}\hat{k}_2 \cdot \ddot{X} \right) -w (\hat{\varepsilon}_1 \cdot \psi ) (\hat{\varepsilon}_2 \cdot \psi) \left(1-w \dot{\phi}\right)\cr
     & \hspace{2cm} - w^2 (\hat{\varepsilon}_1\cdot \psi) (\hat{\varepsilon}_2 \cdot \dot{\psi}) +\frac{w^2}{2} (\hat{\varepsilon}_1 \cdot \hat{\varepsilon}_2)(\dot{\phi}^2 - \ddot{\phi}) + \mathcal{O}(w^3) \Big]\,,
\end{align}
with $\dot{O}(z):= \partial_z O(z)$.
We define $k:=(k_1+k_2)$ and $s_{12}:=(k_1+k_2)^2$. We also used BCFW shift in \eqref{BCFW_mom_shift} to write $\hat{k}_1\cdot \hat{k}_2=k_1\cdot k_2$. All coincident products of operators are understood as normal ordered at $w=0$.
We give a detailed derivation of this expression in appendix \ref{app:Pomeron_vertex}. We now plug the above expression in \eqref{eq:two_point_amp} and do the $w$ integral in the large $z$ regime. After stripping off the deformed massless polarizations, we obtain the integrated vertex operator as follows
\begin{align}\label{eq:int_vertex_pomeron}
    \widehat{\mathcal{V}}^{AB}_{12}(\hat{\varepsilon}_1,\hat{\varepsilon}_2;z,w_3) &=  :\Bigg[ \eta^{AB}\left(1 + \dfrac{i(1-\alpha' s )\dot{\phi}(0)}{\hat{k}_2\cdot \dot{X}(0)}+ \dfrac{i\alpha' s_{12}(1-\alpha' s_{12}) \hat{k}_2\cdot \ddot{X}(0) }{2 (\hat{k}_2 \cdot \dot{X}(0))^2}  + \mathcal{O}(1/z^3) \right)\cr
    &\hspace{2cm} +\psi^A(0)\psi^B(0)\left(\dfrac{\alpha' s_{12}-1}{i\hat{k}_2\cdot \dot{X}(0)}+\mathcal{O}(1/z^2) \right) \Bigg]\mathcal{P}(z): \,,\cr
    & \equiv \widehat{\mathcal{V}}'^{AB}_{12}(z)\,,
\end{align}
in which the $w_3$ dependence is dropped since the $w$ integral in \eqref{eq:two_point_amp} dominates near $\frac{1}{z}$ at large $z$. 
The Pomeron vertex operator $\mathcal{P}(z)$ is defined as (see appendix \ref{app:Pomeron_vertex} for the derivation)
\begin{align}\label{eq:poemron_vertex_massless}
    \mathcal{P}(z) := (-i\hat{k}_2\cdot \dot{X}(0))^{1- \alpha' s_{12}} \Gamma \left[\alpha' s_{12} -1\right] e^{-2\phi(0)} e^{ik \cdot X(0)} \,, \quad \text{for}\,  \quad \mathrm{Re}(1-\alpha' s_{12})<0\,.
\end{align}
The structure of the integrated vertex operator makes the large-$z$ tensor structure manifest. Its leading term is proportional to $\eta^{AB}$, with all remaining dependence contained in an operator without carrying tensor indices of the hard particles. This is the worldsheet manifestation of the enhanced spin-Lorentz symmetry: the helicity remains unchanged for a highly boosted particle moving through a soft background. 
The corrections proportional to $\dot{\phi}$ and $\hat{k}_2 \cdot\ddot{X}$ preserve this structure. In contrast, the normal-ordered bilinear $:\psi^A\psi^B:$ is antisymmetric in $A,B$ and supplies the first symmetry-breaking contribution, suppressed by one power of $z$.

After stripping off the deformed massless polarization vectors let us rewrite the $N-$point amplitude as follows
\begin{align}\label{eq:bcfw_construtible_amp}
    \widehat{\mathcal A}^{h_1,h_2}_N(z) = \hat\varepsilon^{h_1}_{1,A}(z)\hat\varepsilon^{h_2}_{2,B}(z)
    \,\hat{\mathcal A}_N^{AB}(z)\,,
\end{align}
where $(h_1,h_2)$ are the helicities of the deformed massless states and
\begin{align}
    \hat{\mathcal A}_N^{AB}(z) \sim \mathfrak C_{\mathrm{soft}}
    \left[\widehat{\mathcal V}_{12}'{}^{AB}(z)\right]\,.
\end{align}
We define the spectator soft correlator by
\begin{align}
    \mathfrak C_{\mathrm{soft}}[\mathcal O] &:=\int_{\mathcal D'}\prod_{i=3}^{N-2}dw_i\,  \left\langle c(0)\mathcal O\,\prod_{j=3}^{N-2}V_j^{(0)}(k_j,w_j)\,
        cV_{N-1}^{(0)}(k_{N-1},1)\, cV_N^{(0)}(k_N,\infty) \right\rangle\,,
\end{align}
where the reparameterization ghost at the fused insertion has been displayed explicitly.
From large-$z$ expansion of the integrated vertex operator in \eqref{eq:int_vertex_pomeron}, we obtain \cite{Cheung:2010vn}
\begin{align}\label{eq:large_z_massless_type1}
    \widehat{\mathcal A}^{AB}_N(z) \sim{} & z^{-\alpha's_{12}}
    \left[ \eta^{AB}(zc_0+c_1) + \mathrm{O}_1^{AB}  +\frac{\mathrm{O}_2^{AB}}{z} +\mathcal O(z^{-2}) \right]\,,
\end{align}
where the leading term proportional to $\eta^{AB}$ preserves the spin-Lorentz symmetry. The first subleading term that breaks this symmetry is given by an anti-symmetric tensor just as was found in the field theory case and higher subleading terms in $z$ comes with general matrix $\mathrm{O}_2^{AB}$.
From this analysis we now show that any $N-$point amplitude of the form \eqref{eq:bcfw_construtible_amp} is BCFW constructible. To see this, let us now dress $\widehat{\mathcal A}^{AB}_N(z)$ with the deformed polarization vectors to obtain the large $z$ behaviour.

To do this for each helicity configuration, we now restrict the external momenta and gluon polarizations to a four-dimensional subspace \eqref{eq:mom_pol_4d_redux_LR}. The covariant tensor structure in
\eqref{eq:large_z_massless_type1} then allows us to analyze
the shifted amplitude using four-dimensional
spinor helicity variables. We define the
$[1^{h_1}2^{h_2}\rangle$ shift, with $h_{1,2}=\pm1$, by
\begin{align}
    |\hat1\rangle&=|1\rangle,
    &|\hat1]&=|1]-z|2],\\
    |\hat2\rangle&=|2\rangle+z|1\rangle,
    &|\hat2]&=|2]\,,
\end{align}
with $r_{\alpha \dot{\alpha}}= |1\rangle_{\alpha}[2|_{\dot{\alpha}}$.
The deformed massless polarization vectors then takes the following form\cite{Arkani-Hamed:2008bsc}
\begin{align}
 \hat\varepsilon_1^{+,\mu}(z)
    &=-\frac{\sqrt2\big((r^\mu)^*-zk_2^\mu\big)}
    {\langle21\rangle}\,,
    &
    \hat\varepsilon_2^{+,\nu}(z)
    &=-\frac{\sqrt2\,r^\nu}{\langle12\rangle}\,,
    \cr
    \hat\varepsilon_1^{-,\mu}(z)
    &=-\frac{\sqrt2\,r^\mu}{[12]}\,,
    &
    \hat\varepsilon_2^{-,\nu}(z)
    &=-\frac{\sqrt2\big((r^\nu)^*+zk_1^\nu\big)}{[21]}\,.
\end{align}  
Gauge invariance of the deformed amplitude $ \widehat{\mathcal A}_N^{h_1h_2}(z)$
implies the Ward identities
\begin{align}
    \hat k_{1\mu}(z)\,
    \hat{\mathcal A}_N^{\mu\nu}(z)
    \hat\varepsilon_{2\nu}^{h_2}(z)&=0\,,
    \cr
    \hat\varepsilon_{1\mu}^{h_1}(z)\,
    \hat{\mathcal A}_N^{\mu\nu}(z)
    \hat k_{2\nu}(z)&=0\,.
\end{align}
Using $\hat k_1=k_1-z\, r$ and $\hat k_2=k_2+z\, r$, these identities become
\begin{align}\label{eq:Wardid2}
    r_\mu\hat{\mathcal A}_N^{\mu\nu}(z)
    \hat\varepsilon_{2\nu}^{h_2}(z)
    &=
    \frac1z\,k_{1\mu}\hat{\mathcal A}_N^{\mu\nu}(z)
    \hat\varepsilon_{2\nu}^{h_2}(z),
    \cr
    \hat\varepsilon_{1\mu}^{h_1}(z)
    \hat{\mathcal A}_N^{\mu\nu}(z)r_\nu
    &=
    -\frac1z\,\hat\varepsilon_{1\mu}^{h_1}(z)
    \hat{\mathcal A}_N^{\mu\nu}(z)k_{2\nu}.
\end{align}

As an example, let us consider the $[1^+2^+\rangle$ shift. Substituting
the deformed polarizations and using the Ward identities gives
\begin{align}
    \widehat{\mathcal A}_N^{++}(z) = \frac{2}{z \langle12\rangle^2}
    (r^*-z\,k_2)_\mu\, \hat{\mathcal A}_N^{\mu\nu}(z)k_{2\nu}\,.
\end{align}
We then substitute \eqref{eq:large_z_massless_type1} above. The metric terms vanish because $r^*\cdot k_2=0$ and $k_2^2=0$, while anti-symmetry of $\mathrm{O}_{1,\mu \nu}$ implies $ k_{2\mu}\mathrm{O}_1^{\mu\nu}k_{2\nu}=0$.
Thus, we find
\begin{align}
    \widehat{\mathcal A}_N^{++}(z) \sim - \frac{2z^{-\alpha's_{12}-1}}{\langle12\rangle^2}
    \left[ k_{2\mu}\mathrm{O}_2^{\mu\nu}k_{2\nu} -(r^*)_\mu \mathrm{O}_1^{\mu\nu}k_{2\nu}+\mathcal O(z^{-1}) \right] = \mathcal O(z^{-\alpha' s_{12}-1})\,.
\end{align}
Repeating the polarization contractions and Ward-identity analysis
for the remaining helicity choices gives the large $z$ behaviour of the $N-$point amplitude. In Table~\ref{tab:string_large_z}, we tabulate them and compare against the Yang-Mills case \cite{Arkani-Hamed:2008bsc,Cheung:2010vn}.
\begin{table}
    \centering
    \begin{tabular}{c|c|c|c}
        Shift & Yang--Mills & Open superstring
        & Falloff condition \\ \hline
        $[1^-2^+\rangle$
        & $\mathcal O(z^{-1})$
        & $\mathcal O(z^{-\alpha' s_{12}-1})$
        & $\operatorname{Re}\alpha' s_{12} >-1$ \\[2pt]
        $[1^+2^+\rangle$
        & $\mathcal O(z^{-1})$
        & $\mathcal O(z^{-\alpha' s_{12}-1})$
        & $\operatorname{Re}\alpha' s_{12} >-1$ \\[2pt]
        $[1^-2^-\rangle$
        & $\mathcal O(z^{-1})$
        & $\mathcal O(z^{-\alpha' s_{12} -1})$
        & $\operatorname{Re}\alpha' s_{12} >-1$ \\[2pt]
        $[1^+2^-\rangle$
        & $\mathcal O(z^3)$
        & $\mathcal O(z^{3-\alpha' s_{12} })$
        & $\operatorname{Re}\alpha' s_{12} >3$
    \end{tabular}
    \caption{Generic large-$z$ bounds for adjacent massless shifts.}
    \label{tab:string_large_z}
\end{table}
Equivalently, writing the large $z$ behaviour as
$\widehat{\mathcal A}_N^{h_1h_2}(z) =\mathcal O(z^{-\alpha's_{12}-n_{h_1h_2}})$, we find
\begin{align}
    n_{++}=n_{--}=n_{-+}=1, \qquad n_{+-}=-3\,.
\end{align}
Thus, the stringy factor $z^{-\alpha's_{12}}$ allows all four helicity choices to yield valid BCFW recursion in suitable kinematic domains in $\alpha' s_{12}$. In particular, the $[1^+2^-\rangle$ shift, which generically grows as $\mathcal{O}(z^3)$ in Yang-Mills theory,
decays in the case of open superstring amplitude when $\operatorname{Re}(\alpha's_{12})>3$.

The boundary term analysis above concerns the full open superstring amplitude. Its BCFW representation contains a sum over all physical intermediate string states allowed in each factorization channel. Our calculations in subsequent sections select the leading Regge (LR) and subleading Regge (SLR) sectors of this intermediate state sum.

More explicitly, the contribution associated with a Regge trajectory $\mathcal T\in\{\mathrm{LR},\mathrm{SLR}\}$ to four-point open string amplitude is defined by
\begin{align}
    \mathcal A_4^{(\mathcal T)} := \sum_{c}\sum_n \frac{1}{P_c^2+m_n^2} \sum_{\lambda\in\mathcal H_n^{(\mathcal T)}}
    \mathcal A_{3,L}(z_{c,n};\lambda)\, \mathcal A_{3,R}(z_{c,n};\lambda)\,,
\end{align}
where $c$ denotes allowed factorization channels, $\lambda$ denotes the physical states in the Hilbert space $\mathcal H_n^{(\mathcal T)}$ belonging to the trajectory $\mathcal T$ at mass level $n$, and
$z_{c,n}$ is fixed by setting the propagator on-shell $\widehat P_c^{\,2}(z_{c,n}) + m_n^2=0$.
The full recursive representation can therefore be organized as
\begin{align}
    \mathcal A_4 = \mathcal A_{4,\mathrm{LR}} +\mathcal A_{4,\mathrm{SLR}}+\mathcal A_{4, \mathrm{rest}}\,,
\end{align}
with the last term collecting the remaining intermediate states.

Thus, the LR and SLR expressions are defined as selected contributions to the full intermediate-state sum. Their calculation does not require a separate assumption that either
trajectory contribution is independently BCFW constructible with a vanishing boundary term at large $z$.

\section{Three-point amplitudes at subleading Regge}\label{sec:Three-point amplitudes at subleading Regge}

In this section, we construct the three-point amplitudes needed to evaluate the subleading Regge contributions to the four-gluon residue. The corresponding leading Regge amplitudes were obtained in \cite{Schlotterer:2010kk} using
higher-spin vertex operators in open superstring theory. We begin by specifying the three-fermion vertex family in the NS sector considered here and its BRST constraints, and then compute its coupling to two massless gauge bosons.

\subsection{Vertex operator}

We consider the subleading Regge vertex operator in NS sector as follows \cite{Basile:2024uxn}
\begin{align}\label{eq:vertex_subleading_regge_3psi}
    V^{(-1),\mathrm{SLR}}_n(\zeta,k;z) = \frac{i^n}{\sqrt{2\alpha'}^n} \zeta_{A_1\cdots A_n; BCD} :\prod_{i=1}^n  \partial{X}^{A_i} \psi^B \psi^C \psi^D (z) e^{-\phi(z)} e^{i k\cdot X(z)}:\,,
\end{align}
where the higher spin polarization tensor satisfies\footnote{The anti-symmetric bracket is defined as $X^{[a}Y^{b]}:= (X^aY^b -X^b Y^a)$ and $A^{[\mu \nu]}:= (A^{\mu \nu}-A^{\nu \mu})$.}
\begin{align}\label{eq:BRST_subRegge_NS}
\zeta_{A_1\cdots A_n; BCD}
&=\zeta_{(A_1\cdots A_n);[BCD]},
\cr
k^{A_i}\zeta_{A_1\cdots A_n;BCD}&=0,
\qquad i=1,\ldots,n,
\cr
k^B \zeta_{A_1\cdots A_n;BCD }&=0,
\cr
\eta^{A_i A_j}\zeta_{A_1\cdots A_n;BCD }&=0,
\cr
\eta^{A_i B}\zeta_{A_1\cdots A_n; BCD}&=0,
\cr
\zeta_{A_1\cdots A_{i-1}A_{i+1}\cdots A_n;[A_i BCD]}&=0\,.
\end{align}
The last condition is due to the $Q_1$ BRST condition.
The on-shell mass of the subleding-Regge states are: $(\alpha'k^2 +n+1)=0\Rightarrow k^2= - \frac{n+1}{\alpha'}$.

\paragraph{Stress tensor OPE:}

Using the fundamental OPEs collected in appendix \ref{app:OPE_RNS_superstrings}, we derive the stress tensor OPE of the subleading vertex operator with the matter stress tensor given in \eqref{eq:stress_tensor_matter_rns} below and show that $ V^{(-1),\mathrm{SLR}}_n(\zeta,z)$ is indeed a primary operator.

Including the superghost contribution 
\begin{align}
    T_\phi(z)e^{-\phi(w)} \sim \frac{e^{-\phi(w)}}{2(z-w)^2} +\frac{\partial e^{-\phi(w)}}{z-w}\,,
\end{align}
and the matter sectors, we obtain\footnote{For the bosonic sector, we have used the following OPE 
\begin{align}\label{eq:OPE_DXi_V}
\dot{X}_A (z) \prod_{i=1}^{n}\dot{X}_{A_i}(w)e^{ik\cdot X(w)}
\sim \left[-\dfrac{2\alpha'}{(z-w)^2}\sum_{j=1}^{n}\eta_{A A_j}\prod_{i\neq j}\dot{X}_{A_i}(w) -\dfrac{2i\alpha' k_A}{z-w}
\prod_{i=1}^{n}\dot{X}_{A_i}(w)\right]e^{ik\cdot X(w)}\,.
\end{align}
}
\begin{align}\label{eq:SLR_stress_tensor_OPE}
    T(z) V^{(-1),\mathrm{SLR}}_n(\zeta,k;w)
    \sim{} \frac{\mathcal R_4(w)}{(z-w)^4} +\frac{\mathcal R_3(w)}{(z-w)^3} +\frac{n+2+\alpha'k^2}{(z-w)^2} V^{(-1),\mathrm{SLR}}_n(\zeta,k;w)\cr
    +\frac{1}{z-w}\partial V^{(-1),\mathrm{SLR}}_n(\zeta,k;w)\,,
\end{align}
where the total stress tensor is $T(z)= T_m(z)+T_\phi (z)$. The complete symmetry in the $\{\mu_i\}$-indices of the massive polarization gives
\begin{align}
    \mathcal R_4(w)  ={}& -i^n\frac{\alpha'n(n-1)}{\sqrt{2\alpha'}^n}\,
    \eta^{EF}  \zeta_{EF A_1\cdots A_{n-2};BCD}
    : \prod_{i=1}^{n-2}\partial X^{A_i}\,\psi^B \psi^C \psi^D
    e^{-\phi}e^{ik\cdot X}:(w)\,,
    \\
    \mathcal R_3(w) ={}& -i^{n+1}\frac{2\alpha'n}{\sqrt{2\alpha'}^n}\,
    k^E \zeta_{ E A_1\cdots A_{n-1};BCD }: \prod_{i=1}^{n-1}\partial X^{A_i}\, \psi^B \psi^C \psi^D e^{-\phi}e^{ik\cdot X} :(w)\,.
\end{align}
To remove these higher order poles in the OPE, we therefore impose the tracelessness and transversality constraints
\begin{align}
    \eta^{EF } \zeta_{EF A_1\cdots A_{n-2};BCD} =0\,, \qquad 
    k^E \zeta_{E A_1\cdots A_{n-1};BCD}=0\,.
\end{align}
Thus $V^{(-1),\mathrm{SLR}}_n(\zeta,k;w)$ is a conformal primary with conformal weight 
\begin{align}
    h = n+\frac32+\frac12+\alpha'k^2 = n+2+\alpha'k^2\,.
\end{align}
On the mass shell $\alpha'k^2=-(n+1)$, we have $h=1$, and therefore
\begin{align}
    T(z)V^{(-1),\mathrm{SLR}}_n(\zeta,k;w) \sim \frac{1}{(z-w)^2} V^{(-1),\mathrm{SLR}}_n(\zeta,k;w) +\frac{1}{z-w}\partial V^{(-1),\mathrm{SLR}}_n(\zeta,k;w)\,.
\end{align}
Next we check the BRST closure.

\paragraph{Supercurrent OPE:}

We use the RNS supercurrent defined in \eqref{eq:supercurrent_def_rns} to derive its OPE with the subleading Regge vertex operator in \eqref{eq:vertex_subleading_regge_3psi}. We obtain
\begin{align}\label{eq:SLR_supercurrent_OPE}
    G(z)V^{(-1),\mathrm{SLR}}_n(\zeta,k;w) \sim \frac{\mathcal S_3(w)}{(z-w)^3}
    +\frac{\mathcal S_2(w)}{(z-w)^2} +\frac{\mathcal S_1(w)}{z-w}\,.
\end{align}
We used the supercurrent OPEs \eqref{eq:supercurrent_opes} in appendix \ref{app:OPE_RNS_superstrings}.

The third-order pole arises from simultaneously contracting
$\partial X^A (z)$ with a $\partial X^{A_i}(w)$ and $\psi_A (z)$ with
one of the three fermions in the vertex operator. We get
\begin{align}
    \mathcal S_3(w) ={}& -\frac{6i^{n+1}\alpha'n }{4\sqrt{2\alpha'}\,(2\alpha')^{n/2}}\, \eta^{E B} \zeta_{E A_1\cdots A_{n-1}; BCD}
    :\prod_{i=1}^{n-1}\partial X^{A_i}\, \psi^C \psi^D e^{-\phi}e^{ik\cdot X}:(w)\,.
\end{align}
This pole vanishes upon imposing the mixed trace condition
\begin{align}\label{eq:mixed trace}
    \eta^{EB} \zeta_{E A_1\cdots A_{n-1};BCD }=0\,.
\end{align}
The second-order pole has two contributions:
\begin{align}
    \mathcal S_2(w) = \frac{i^n}{4\sqrt{2\alpha'}\,(2\alpha')^{n/2}}:
    \Bigg[-2i\alpha'n\,
        \zeta_{E A_1\cdots A_{n-1}; BCD} \prod_{i=1}^{n-1}\partial X^{A_i}\,
        \psi^E \psi^B \psi^C \psi^D \cr
        +6\alpha'k^B \zeta_{A_1\cdots A_n;BCD} \prod_{i=1}^{n}\partial X^{A_i}\, \psi^C \psi^D \Bigg]e^{-\phi}e^{ik\cdot X}: (w)\,.
\end{align}
The first term comes from contracting $\partial X(z)$ with a bosonic matter in the vertex operator while leaving $\psi(z)$ uncontracted. The second term arises from simultaneously contracting $\partial X(z)$ with the plane wave factor and $\psi(z)$ with one of the fermions in the vertex operator.
Since the remaining four fermions are totally antisymmetric we impose the following constraints on the massive polarization $\zeta$
\begin{align}\label{eq:fermionic transversality_yong}
    \zeta_{A_1\cdots A_{n-1}[A_n;BCD]}=0\,,\qquad  k^B \zeta_{ A_1\cdots A_n; BCD}=0\,.
\end{align}
Lastly, the simple-pole coefficient is
\begin{align}
    \mathcal S_1 (w) ={}&
    \frac{i^n}{4\sqrt{2\alpha'}\,(2\alpha')^{n/2}}:
    \Bigg[-2i\alpha'n\,
        \zeta_{E A_1\cdots A_{n-1};BCD } \prod_{i=1}^{n-1}\partial X^{A_i}\,
        \partial\psi^E \psi^B \psi^C \psi^D \cr
    &\hspace{2cm} +3i\, \zeta_{ A_1\cdots A_n; BCD}
        \prod_{i=1}^{n}\partial X^{A_i}\,\partial X^B \psi^C \psi^D \cr
    &\hspace{2cm}
        +2\alpha'\,\zeta_{ A_1\cdots A_n;BCD}
        \prod_{i=1}^{n}\partial X^{A_i}\,(k\cdot \psi)\psi^B \psi^C \psi^D \Bigg]e^{-\phi}e^{ik\cdot X}:(w)\,.
\end{align}
Consequently, imposing the set of constraints in \eqref{eq:mixed trace} and \eqref{eq:fermionic transversality_yong} removes all poles except the simple one
\begin{align}
    G(z)V^{(-1),\mathrm{SLR}}_n(\zeta,k;w) \sim \frac{\mathcal S_1(w)}{z-w}\,.
\end{align}
These are the additional supercurrent constraints, to be combined with the constraints obtained from the stress-tensor OPE. We listed them in \eqref{eq:BRST_subRegge_NS}.

\subsection{Three-point amplitude}

We consider the three-point amplitude with one massive higher spin state at sub-leading Regge trajectory, two massless gluon states in $0$ and in canonical $(-1)$ superghost picture-both at leading Regge trajectory. Recall that, the gluon vertex operator in $(-1)$ picture is \eqref{eq:gluon_vertex_op_can_pic} and the ghost picture $0$ vertex operator at leading Regge trajectory is given by
\begin{align}
    V_{\gamma}^{(0)} (\varepsilon_1,z_1) =  \varepsilon_{1 B} \left[ (k_1\cdot \psi(z_1))\psi^B(z_1) +\frac{i}{2\alpha'}  \dot{X}^B (z_1) \right] e^{i k_1\cdot X(z_1)}\,.
\end{align}
Using these vertex operators, the amplitude in ten dimensions is derived as follows
\begin{align}\label{eq:three_pt_sub_Regge}
    \mathcal{A}(1_{\mathrm{LR}}^{h_1}, 2_{\mathrm{LR}}^{h_2}, \mathbf{3}_{\mathrm{SLR}}^n) &= \left\langle  c V_{\gamma}^{(0)} (\varepsilon^{h_1}_1,z_1) c V_{\gamma}^{(-1)}(\varepsilon^{h_2}_2,z_2) cV^{(1,n)}_{(-1)}(\zeta_3,z_3) \right\rangle \cr
    &= \langle c(z_1)c(z_2)c(z_3)\rangle \langle e^{-\phi(z_2)} e^{-\phi(z_3)}\rangle \mathcal{A}_3 \,.
\end{align}
We have omitted the Chan-Paton factors as we consider only colour ordered amplitude throughout this paper.
We denote the massive state with bold face notation, $h_i= \pm 1 $ denote gluon helicities and define $\mathcal{A}_3$ as
\begin{align}
    \mathcal{A}_3 := \frac{i^n}{\sqrt{2\alpha'}^n} \varepsilon^{h_1}_{1,A_1} \varepsilon^{h_2}_{2,A_2} \zeta_{3,C_1\cdots C_n; BCD} \Big\langle \Big[ (k_1\cdot \psi_1)\psi_1^{A_1} +\dfrac{i}{2\alpha'}\dot{X}_1^{A_1}  \Big] e^{i k_1\cdot X_1} \,,  \psi_2^{A_2} e^{ik_2\cdot X_2}\,,\cr
    \prod_{i=1}^n \dot{X}_3^{C_i} \psi_3^B \psi_3^C \psi_3^D e^{ik_3\cdot X_3} \Big\rangle \,.
\end{align}
Here, we denote all the RNS fields $\mathcal{O}(z_i)\equiv \mathcal{O}_i$ where the subscript $i$ denotes the location of the RNS operator $\mathcal{O}$ on the disk.
The Bosonic part of the picture-$0$ gluon vertex operator vanishes trivially since 
\begin{align}
    \langle \psi_2^{A_2} \psi_3^B \psi_3^C \psi_3^D \rangle =0\,,
\end{align}
and we obtain $\mathcal{A}_3$ straightforwardly
\begin{align}
    \mathcal{A}_3= - (-1)^n (2\alpha')^{n/2} (\zeta_{3,BCD}\cdot k_2^n) \left(\varepsilon_2^{B,h_2} k_1^{[D}\varepsilon_1^{C],h_1} - \varepsilon_2^{C , h_2} k_1^{[D}\varepsilon_1^{B],h_1} + \varepsilon_2^{D ,h_2} k_1^{[C}\varepsilon_1^{B],h_1} \right) \cr
    \times \frac{z_{12}^n}{z_{32}^{n+1} z_{31}^{n+2}}\prod_{a<b}^3 z_{ab}^{2\alpha' k_a\cdot k_b}\,,
\end{align}
using the shorthand notation of \cite{Schlotterer:2010kk} to denote
\begin{align}
    (\zeta_{BCD}\cdot k^n) \equiv \zeta_{A_1 \cdots A_n,BCD} k^{A_1}\cdots k^{A_n}\,.
\end{align}
Including the ghost and superghost sectors
\begin{align}
    \langle c(z_1)c(z_2)c(z_3)\rangle= z_{12}z_{13}z_{23}\,,\qquad
    \langle e^{-\phi(z_2)}e^{-\phi(z_3)}\rangle=\frac{1}{z_{23}}\,,
\end{align}
we find the dependence on the coordinates of string worldsheet $z_i$ cancels out in the amplitude as expected
\begin{align}
    \frac{z_{12}^{\,n+1}} {z_{31}^{\,n+1}z_{32}^{\,n+1}}
    \prod_{a<b}^{3}z_{ab}^{2\alpha'k_a\cdot k_b} =1\,,
\end{align}
Thus, we find the three-point amplitude as
\begin{align}\label{eq:three-point-subregge}
     \mathcal{A}_3(1_{\mathrm{LR}}^{h_1}, 2_{\mathrm{LR}}^{h_2}, \mathbf{3}_{\mathrm{SLR}}^n) &= (-1)^n (2\alpha')^{n/2} (\zeta_{3,BCD}\cdot k_2^n) \left[\varepsilon_2^{B} k_1^{[D}\varepsilon_1^{C]} - \varepsilon_2^C k_1^{[D}\varepsilon_1^{B]} + \varepsilon_2^D k_1^{[C}\varepsilon_1^{B]} \right]\,.
\end{align}
To avoid clutter, we suppressed $h_i$'s on the massless polarization vectors.
To check the mass dimension, we include the Yang-Mills coupling whose mass dimension in $D$ dimensions is $[g_{\mathrm{YM}}]_m = \left( 2- \frac{D}{2} \right)$. Together with the fact that all polarizations are dimensionless in any dimensions and $[\alpha']_m =-2$, we find the mass dimension of the amplitude as $[g_{\mathrm{YM}}\mathcal{A}(1^{h_1}, 2^{h_2}, \mathbf{3}^n)]_m = -2$, as expected in ten dimensions\footnote{In $D$ dimension, the mass dimension of an $n-$point amplitude is $[\mathcal{A}_n]_m = D- \frac{n(D-2)}{2}$.}.
As a simple consistency check, we find this amplitude obeys
\begin{align}
      \mathcal{A}(1_{\mathrm{LR}}^{h_1}, 2_{\mathrm{LR}}^{h_2}, \mathbf{3}_{\mathrm{SLR}}^n) \Big|_{\varepsilon_1\to k_1}=0\,, \qquad  \mathcal{A}(1_{\mathrm{LR}}^{h_1}, 2_{\mathrm{LR}}^{h_2}, \mathbf{3}_{\mathrm{SLR}}^n) \Big|_{\varepsilon_2\to k_2}=0\,,
\end{align}
as a consequence of the Ward identities satisfied by the external massless states. The first identity is straightforward from the anti-symmetric structure in ($k_1,\varepsilon_1$) in \eqref{eq:three-point-subregge}, while to get the second Ward identity, we make use of the last equation in the set of BRST constraints \eqref{eq:BRST_subRegge_NS} satisfied by the massive polarization $\zeta_3$.

\section{Four-gluon residue at leading Regge} \label{sec:Four gluon residue at leading Regge}

In this section, we compute the leading Regge contribution to the residue of color-ordered four-gluon amplitude
$\mathcal{A}_4(1^+,2^+,3^-,4^-)$. We begin by deriving the four-dimensional amplitude from the known
ten-dimensional three-point amplitudes of
\cite{Schlotterer:2010kk} and expressing them in spinor helicity variables. We then glue these amplitudes using BCFW recursion, selecting leading Regge states among the intermediate-state sum. The resulting little group contractions yield a compact expression for the residues in terms of Jacobi polynomials. Finally, we examine the Regge limit of the corresponding mass-level sum.

\subsection{Three-point amplitudes and its 4d reduction}

The ten-dimensional three-point amplitude involving two gluons and a massive higher spin state at leading Regge trajectory was obtained in eq. (3.8) of \cite{Schlotterer:2010kk}. In our notation, it is given by
\begin{align}\label{eq:three-point-leading_regge}
\mathcal A_3 (1_{\mathrm{LR}},2_{\mathrm{LR}},\mathbf3^n_{\mathrm{LR}})
= \frac{(2\alpha')^{\frac{n+1}{2}}}{\sqrt{\alpha'}} \Bigg[\frac{n}{2\alpha'}\, \varepsilon_1^A \varepsilon_2^B \left(\xi_{\vec k_3,AB}\cdot k_1^{n-1}\right) +(\varepsilon_2\cdot k_3)\varepsilon_1^A
\left(\xi_{\vec k_3,A}\cdot k_1^n\right) \cr
+(\varepsilon_1\cdot k_2)\varepsilon_2^A
\left(\xi_{\vec k_3,A}\cdot k_1^n\right) + (\varepsilon_1\cdot\varepsilon_2)
\left(\xi_{\vec k_3}\cdot k_1^{n+1}\right)\Bigg]\,.
\end{align}
Following \cite{Schlotterer:2010kk}, we use the shorthand
\begin{align}
\xi_{A_1\cdots A_j}\cdot k^{s-j} := \xi_{A_1\cdots A_j A_{j+1}\cdots A_s}
k^{A_{j+1}}\cdots k^{A_s}\,.
\end{align}
At mass level $n\geq1$, the massive state has spin $J_n = n+1$ and mass $\alpha' m_n^2=n $. Its polarization tensor is totally symmetric,
\begin{align}
\xi_{\vec k,A_1\cdots A_{n+1}} = \xi_{\vec k,(A_1\cdots A_{n+1})}\,,
\end{align}
and satisfies the transversality and tracelessness conditions due to BRST invariance
\begin{align}\label{eq:10d_brst_LR}
k^{A_i}\xi_{\vec k,A_1\cdots A_{n+1}}
&=0,
&& i=1,\ldots,n+1\,,
\\
\eta^{A_i A_j}\xi_{\vec k,A_1\cdots A_{n+1}}&=0,
&& 1\leq i<j\leq n+1\,.
\end{align}

We obtain the four-dimensional amplitude from \eqref{eq:three-point-leading_regge} as follows. We parametrize all momenta and polarization tensors within a four-dimensional subspace 
\begin{align}\label{eq:mom_pol_4d_redux_LR}
    k^A = (k^\mu,\vec 0)\,, \qquad \varepsilon_{\vec k}^{A} = (\varepsilon_{\vec k}^{\mu},\vec 0)\,,
\end{align}
where $(\mu ,\nu ,\ldots)=(0,\ldots,3)$- labels the 4d indices.
We define the four-dimensional spin-$n$ polarization \cite{Guevara:2018wpp, Chung:2018kqs, Johansson:2019dnu}
\begin{align}\label{eq:massive_polarization_spin_n}
    E_{\vec k,c_1\cdots c_{2n}}^{\rho_1\cdots\rho_n} := \frac{1}{n!} \mathcal S_{c_1\cdots c_{2n}}
    \left[ \varepsilon_{\vec k,c_{1}c_{2}}^{(\rho_1} \cdots \varepsilon_{\vec k,c_{2n-1}c_{2n}}^{\rho_n)}\right]\,,
\end{align}
where $\mathcal S$ denotes complete symmetrization over the $2n$ SU(2) little-group indices $\{c_i\}$ without any additional normalization.
This selects the spin-$n$ representation of SU(2) group. The resulting tensor $E$ is symmetric and traceless in its Lorentz indices and transverse w.r.t the 4d momentum $k^\mu$
\begin{align}\label{eq:E_tensor_transverse_traceless}
    k_{\rho_i}E_{\vec k,\mathbf c}^{\rho_1\cdots\rho_n} =0\,, \qquad \eta_{\rho_i\rho_j}
    E_{\vec k,\mathbf c}^{\rho_1\cdots\rho_n}  = 0 \qquad \forall i\neq j\,,
\end{align}
where $\mathbf c=(c_1,\ldots,c_{2n})$. 
Transversality follows directly from the orthogonality relation : $(k \cdot \varepsilon_{\vec{k},ab})=0$ and the tracelessness condition can be realized in spinor helicity basis as follows
\begin{align}
\eta_{\mu\nu}E^{\mu\nu}_{\vec{k}, c_1\cdots c_4} = 2 \varepsilon_{\vec{k}, c_1c_2} \cdot \varepsilon_{\vec{k}, c_3c_4} \sim \mathcal{S}_{c_1\cdots c_{4}}\Big[\langle k_{c_1} k_{c_3}\rangle [k_{c_2} k_{c_4}] \Big] =0\,.    
\end{align}

Throughout this paper, we use the massive spinor helicity formalism of \cite{Arkani-Hamed:2017jhn} and the conventions and definitions of \cite{Ochirov:2018uyq}.
The Dirac equations satisfied by the massive spinor helicity variables are
\begin{align}\label{eq:Dirac_eqn_spinors}
\hat P|\hat P^a] &= m_P|\hat P^a\rangle\,, \quad &&\hat P|\hat P^a\rangle = m_P|\hat P^a]\,,\cr
\langle\hat P^a|\hat P &=-m_P[\hat P^a|\,, \quad &&[\hat P^a|\hat P = -m_P\langle\hat P^a|\,.
\end{align}
The massless polarization vectors are defined as
\begin{align}\label{eq:massless_polarizations_spinorhelicity}
    \varepsilon^{+,\mu}_{\vec{p}}:=\frac{\langle q|\sigma^\mu |p]}{\sqrt{2}\langle q p \rangle}\,, \qquad \qquad \varepsilon^{-,\mu}_{\vec{p}}:=\frac{\langle p|\sigma^\mu |q]}{\sqrt{2} [p q]}\,, 
\end{align}
for the positive and negative helicity states with reference spinor $q$ whereas the spin-1 massive polarization is defined as \cite{Guevara:2018wpp}
\begin{align}\label{eq:massive_polarizations_spinorhelicity}
	 \varepsilon_{\vec{p},ab}^{\mu} := \frac{i}{\sqrt{2}m} \langle p_{(a}|\sigma^\mu|p_{b)}]\,,
\end{align}
satisfying $p_\mu \varepsilon_{\vec{p},ab}^{\mu} =0$.
Therefore, we take the 4d completion of the massive higher spin polarization in 10d as
\begin{align}
    \xi_{\vec k}^{\rho_1\cdots \rho_{n}}= E_{\vec k,c_1\cdots c_{2n}}^{\rho_1\cdots\rho_n}\,,
\end{align}
satisfying the BRST conditions in \eqref{eq:10d_brst_LR} and we set all other components of $\xi$ to zero. 

Using this representation of the massive higher spin polarization, the three-point amplitude in 4d is given by
\begin{align}\label{eq:three-point-LR-4d}
    \mathcal A_3 (1^{h_1}_{\mathrm{LR}},2^{h_2}_{\mathrm{LR}},\mathbf3^n_{\mathrm{LR}}) = \dfrac{(2\alpha')^{\frac{n+1}{2}}}{\sqrt{\alpha'}n!} \left[(\varepsilon_3 \cdot k_1)\chi \right]^{\odot 2(n+1)} \,,
\end{align}
with
\begin{align}\label{eq:core_tensor_3pt_LR}
    \chi := \Bigg[ \frac{n}{2\alpha'} (\varepsilon_2^{h_2} \cdot \varepsilon_{3}) (\varepsilon_1^{h_1} \cdot \varepsilon_{3}) + \left\{ -(\varepsilon_2^{h_2} \cdot k_1) (\varepsilon_1^{h_1} \cdot \varepsilon_{3})+ (\varepsilon_2^{h_2} \cdot \varepsilon_{3}) (\varepsilon_1^{h_1} \cdot k_2) \right\}(\varepsilon_{3}\cdot k_1)\cr
    + (\varepsilon_2^{h_2} \cdot \varepsilon_1^{h_1} )(\varepsilon_{3}\cdot k_1)^2 \Bigg]\,.
\end{align}
The SU(2) indices corresponding to the intermediate spin $s=(n+1)$ massive states left implicit in the definition of massive spin-1 polarization.
Here $\odot 2(n+1)$ denotes normalized complete symmetrization over all such indices of the combined product\cite{Aoude:2021oqj}.

\subsection{BCFW shift and four-point residue}

We now consider the contribution to the four-point residue of the colour-ordered amplitude $\mathcal A_4(1^+,2^+,3^-,4^-)$ at leading Regge trajectory.
We use the $[1^+\,2^+\rangle$ shift\footnote{The same helicity sector of the 4d three-point amplitudes in \eqref{eq:three-point-LR-4d} vanish \cite{Feng:2011qc}, see appendix \ref{app:same_helicity_LR}. This particular shift uses non-zero opposite helcity sector amplitudes.}, which uses only the $s_{14}$ factorization channel. The complexified momenta are
\begin{align}\label{eq:BCFW_mom_shift}
\hk_1^\mu=k_1^\mu-zr^\mu,
\qquad
\hk_2^\mu=k_2^\mu+zr^\mu.
\end{align}
In spinor helicity variables, this deformation is realized by
\begin{align}
|\hat1]&=|1]-z|2],
&
|\hat2\rangle&=|2\rangle+z|1\rangle,
\\
|\hat1\rangle&=|1\rangle,
&
|\hat2]&=|2],
\end{align}
with $r_{\alpha\dot\alpha}=|1\rangle_\alpha[2|_{\dot\alpha}$.

From momentum conservation at each of the sub-amplitudes, the internal momentum is given by $\hat P_n(z) = -(k_4+\hat k_1(z)) = (\hat k_2(z)+k_3)$. All shifted spinor brackets are evaluated at $z=z_{14}$. At mass level $n$, we determine this by setting the shifted propagator on shell
\begin{align}\label{eq:LR_z14_pole}
 (\hat{s}_{14} + m_n^2)_{z=z_{14}}=0\, \Rightarrow z_{14,n} = \frac{(2k_1\cdot k_4 + m_n^2)}{2r\cdot k_4}\,, \qquad m_n^2=\frac{n}{\alpha'}\,.
\end{align}
We denote the contribution of the leading Regge states to the four-point residue as
\begin{align}\label{eq:LR_gluing_s14_channel}
\mathcal R_n^{\mathrm{LR}}(1^+,2^+,3^-,4^-) :={}& \left.\mathcal A_3 (4^-_{\mathrm{LR}},\hat1^+_{\mathrm{LR}},
\hat{\mathbf P}_{n,\mathrm{LR}})
\mathcal A_3 (\hat2^+_{\mathrm{LR}},3^-_{\mathrm{LR}},-\hat{\mathbf P}_{n,\mathrm{LR}}) \right|_{z=z_{14,n}},
\end{align}
where contraction of the internal SU(2) little group indices
implements the intermediate-state sum. Therefore, the corresponding contribution to the BCFW recursion is
\begin{align}\label{eq:LR_bcfw_recursion}
A^{\mathrm{LR}}(1^+,2^+,3^-,4^-) =\frac{1}{s_{14}}\mathcal R_0^{\mathrm{LR}}(1^+,2^+,3^-,4^-) + \sum_{n=1}^{\infty} \frac{\mathcal R_n^{\mathrm{LR}}(1^+,2^+,3^-,4^-)}{s_{14}+m_n^2}\,.
\end{align}

We first consider the $n=0$ level, corresponding to the massless exchange. Its residue is obtained by gluing the standard three-gluon MHV and anti-MHV amplitudes and summing over the two intermediate helicities. The amplitudes are
\begin{align}
     \mathcal{A}_3(4^-,\hat{1}^+,\hat{P}_{0}^+) &= \frac{[\hat{1}\hat{P}]^3}{[4\hat{1}][4\hat{P}]}  \,, \quad  \mathcal{A}_3(4^-,\hat{1}^+,\hat{P}_{0}^-) = \frac{\langle 4\hat{P}\rangle^3}{\langle 4 1\rangle \langle 1\hat{P}\rangle} \\
      \mathcal{A}_3(\hat{2}^+,3^-,-\hat{P}_{0}^+) &= \frac{[\hat{P}2]^3}{[23][3\hat{P}]} \,, \quad  \mathcal{A}_3(\hat{2}^+,3^-,-\hat{P}_{0}^-) = \frac{\langle 3\hat{P}\rangle^3}{\langle \hat{2}3\rangle \langle \hat{2} \hat{P}\rangle}\,.
\end{align}
At the $\hat{s}_{14}=0$ factorization channel, both $\mathcal{A}_3(4^-,\hat{1}^+,\hat{P}_{0}^+)$ and $\mathcal{A}_3(\hat{2}^+,3^-,-\hat{P}_{0}^-)$ vanishes and we obtain the massless residue as
\begin{align}\label{eq:R0_LR_amp}
    \mathcal{R}_0^{\mathrm{LR}}(1^+,2^+,3^-,4^-)&= \mathcal{A}_3(4^-,\hat{1}^+,\hat{P}_{0}^-)\, \mathcal{A}_3(\hat{2}^+,3^-,-\hat{P}_{0}^+) =\frac{1}{s_{12}}[12]^2 \langle 34\rangle^2 \,.
\end{align}

We next consider the massive leading Regge states at levels $n\geq1$. We organize each three-point amplitude into $n-1$ symmetric rank-2 tail tensors and a symmetric rank-4 core tensor
\begin{align}
     \mathcal{A}_3(4^-_{\mathrm{LR}},\hat{1}^+_{\mathrm{LR}},\hat{\mathbf{P}}_{n,\mathrm{LR}}) &= \frac{(2\alpha')^{\frac{n+1}{2}}}{\sqrt{\alpha'}n!} \left( \frac{i}{\sqrt{2}m_n}\right)^{n-1}  (\mathbb{X}_4 \, \chi_L)^{\odot 2(n+1)} \,, \cr
     \mathcal{A}_3(\hat{2}^+_{\mathrm{LR}},3^-_{\mathrm{LR}},-\hat{\mathbf{P}}_{n,\mathrm{LR}}) &= \frac{(2\alpha')^{\frac{n+1}{2}}}{\sqrt{\alpha'}n!} \left( \frac{-i}{\sqrt{2}m_n}\right)^{n-1}  (\mathbb{X}_2 \, \chi_R )^{\odot 2(n+1)} \,.
\end{align}
Here we define the \textit{tail} tensors as
\begin{align}
\mathbb{X}_4^{ab}:= \langle \hat{P}^{(a}|k_4|\hat{P}^{b)}] \,, \qquad     \mathbb{X}_2^{ab}:= \langle \hat{P}^{(a}|\hat{k}_2|\hat{P}^{b)}]\,,
\end{align}
 and the \textit{core} tensors as
\begin{align}\label{eq:LR_core_tensors}
    \chi_L^{\odot 4} := &\Bigg[ \frac{n}{2\alpha'} (\he_1^+ \cdot \hat{\varepsilon}_{\vec{P}}) (\varepsilon_4^- \cdot \hat{\varepsilon}_{\vec{P}}) + \{ - (\he_1^+ \cdot k_4) (\varepsilon_4^- \cdot \hat{\varepsilon}_{\vec{P}}) \cr
    &\hspace{2cm} + (\he_1^+ \cdot \hat{\varepsilon}_{\vec{P}}) (\varepsilon_4^- \cdot \hk_1) \}(\varepsilon_{\vec{P}}\cdot k_4)
    + (\he_1^+ \cdot \varepsilon_4^- ) (\varepsilon_{\vec{P}}\cdot k_4)^2 \Bigg]\,,
\end{align}
 \begin{align}
    \chi_R^{\odot 4} :=& \Bigg[ \frac{n}{2\alpha'} (\he_2^{+}\cdot \hat{\varepsilon}_{-\Vec{P}}) (\varepsilon_3^{-}\cdot \hat{\varepsilon}_{-\Vec{P}})
     - (\varepsilon_{3}^- \cdot \hat{k}_2) (\he_2^{+}\cdot \hat{\varepsilon}_{-\Vec{P}}) (\hat{\varepsilon}_{-\Vec{P}}\cdot \hk_2) \cr
     &\hspace{2cm} +(\he_2^+ \cdot k_3)(\varepsilon_3^{-} \cdot \hat{\varepsilon}_{-\Vec{P}}) ( \hat{\varepsilon}_{-\Vec{P}} \cdot  \hk_2) 
    + (\varepsilon_3^- \cdot \he_2^+) ( \hat{\varepsilon}_{-\Vec{P}} \cdot \hk_2)^{2} \Bigg]\,,
\end{align}
where the SU(2) indices associated to the massive spin-1 polarization $\varepsilon^\mu_{\vec{P}}$ are left implicit, we restore them while gluing these three-point amplitude to obtain the residue.
The combined product is defined using $\odot n$ as follows 
\begin{align}
(\mathbb X_4\chi_L)^{\odot 2(n+1)} :=\frac{1}{\mathcal N_n}
\sum_{(i,j)\in\mathcal{S}'_n}
\left[\prod_{r=1}^{n-1}\mathbb X_4^{a_{i_r}a_{j_r}} \right]
\chi_L^{a_{i_n}a_{i_{n+1}}a_{j_n}a_{j_{n+1}}}\,,
\end{align}
where $\mathcal{S}'_n$ enumerates distinct assignments of four SU(2) indices from of set of $2(n+1)$ indices to the symmetric core and unordered pairings of the remaining indices into the identical symmetric tails.
The normalization included in this definition is
\begin{align}
\mathcal N_n = \binom{2n+2}{4}(2n-3)!! = \frac{(2n+2)!}{4!\,2^{n-1}(n-1)!}\,.
\end{align}
with $(-1)!!=1$.
The right product is defined analogously.
Together with the right sub-amplitude, we obtain the four-point residue at mass level $n\geq 1$ as 
\begin{align}\label{eq:LR_massive_levels}
  \mathcal R_{n\geq 1}^{\mathrm{LR}}(1^+,2^+,3^-,4^-) =\frac{1}{\alpha'(n!)^2} \frac{ (2\alpha')^{n+1}}{(2m_n^2)^{n-1}}\mathcal{F}_n\,,
\end{align}
with
\begin{align}
    \mathcal{F}_n := (\mathbb{X}_4\chi_L)^{\odot 2(n+1)} (\mathbb{X}_2\chi_R)_{\odot 2(n+1)}\,.
\end{align}
The super/sub-scripts $\odot 2(n+1)$ denotes the position of the symmetric SU(2) indices.

\subsection{Evaluation of $\mathcal{F}_n$}

We now evaluate the SU(2) little-group contraction $\mathcal F_n$
that determines the massive leading Regge residues. We first choose the null reference momenta entering the massless polarization vectors as
\begin{align}\label{eq:LR_ref_spinors}
r_1=r_4=k_3\,, \qquad r_2=r_3=k_4\,,
\end{align}
and express the tail and core tensors in a common chiral basis. Their factorized forms then allow us to organize the normalized symmetrization and perform the
contraction between the two three-point sub-amplitudes.

\paragraph{Chiral factorization of the tail and core tensors :}

We introduce the chiral spinor helicity brackets
\begin{align}\label{eq:chiral_spinor_bracket_P}
u^a&:=\langle4\hat P^a\rangle,
&
w^a&:=\langle1\hat P^a\rangle,
\\
v^a&:=\langle3\hat P^a\rangle,
&
t^a&:=\langle\hat2\hat P^a\rangle.
\end{align}
We first derive the chiral representation of the tails.
Using the massive Dirac equations\footnote{Note $\hat P=-(\hat k_1+k_4)=\hat k_2+k_3$. We suppress the level dependence on $\hat{P}$ here.} given in \eqref{eq:Dirac_eqn_spinors}, we convert all $[q\hat P^a]$ as follows
\begin{align}
[q\hat P^a] = \frac{1}{m_n}\langle\hat P^a|\hat P|q]\,,
\end{align}
for any massless spinor $q$. Moreover on the factorization pole at $z=z_{14,n}$ given in \eqref{eq:LR_z14_pole}, the shifted spinor helicity brackets evaluates to
\begin{align}\label{eq:LR_shifted_brackets}
[\hat14]=\frac{m_n^2}{\langle41\rangle}\,, \qquad [32]=-\frac{m_n^2}{\langle3\hat2\rangle}\,.
\end{align}
Consequently, we write the tail tensors as
\begin{align}
\mathbb X_4^{ab} = -\frac{m_n}{\langle41\rangle}u^{(a}w^{b)}\,, \qquad
\mathbb X_2^{ab}= -\frac{m_n}{\langle3\hat2\rangle}v^{(a}t^{b)}\,.
\end{align}
Here we used $\langle\hat P^a|k_4|4]=0$ and $\langle\hat P^a|\hat k_2|2]=0$,
together with \eqref{eq:LR_shifted_brackets}.

We now simplify the left core. Let $\chi_L := \sum_{j=1}^4 \chi_{L,j}$ in its covariant definition \eqref{eq:LR_core_tensors}, in their original order. In the following intermediate expressions, $\mathcal S$ denotes
normalized symmetrization over the four indices; for example,
\begin{align}
\mathcal S[u^3w]\equiv \mathcal S[u^au^bu^cw^d]\,,\qquad \mathcal S[u^2 w^2]\equiv \mathcal S[u^a u^b w^cw^d]\,.
\end{align}
Note, Schouten identity and momentum conservation give
\begin{align}\label{eq:int_Schouten_X_L_comp}
v^a =
\frac{\langle31\rangle u^a+\langle43\rangle w^a}{\langle41\rangle}\,,\qquad
t^a = -\frac{\langle12\rangle}{\langle41\rangle}
\left(u^a+\frac{[\hat13]}{[43]}w^a \right)\,.
\end{align}
Using this, we rewrite each $\chi_{L,j}$. For example,
\begin{align}
\chi_{L,1} = -\frac{1}{2\langle31\rangle\langle12\rangle} \mathcal S[u^2vt]\,,
\end{align}
where we used $\alpha'm_n^2=n$. Substituting \eqref{eq:int_Schouten_X_L_comp} and \eqref{eq:LR_shifted_brackets}, and similarly simplifying the remaining terms yields
\begin{align}
\chi_{L,1} &= \frac{u^a u^b u^c u^d}{2\langle41\rangle^2}
+\beta_L\,\mathcal S[u^3w] +\gamma_L\,\mathcal S[u^2w^2]\,,
\\
\chi_{L,2}+\chi_{L,3} &= -\beta_L\,\mathcal S[u^3w] -2\gamma_L\,\mathcal S[u^2w^2]\,,
\\
\chi_{L,4} &= \gamma_L\,\mathcal S[u^2w^2]\,,
\end{align}
with
\begin{align}
\beta_L := \frac{\langle31\rangle[\hat13]+\langle43\rangle[43]}
{2\langle31\rangle[43]\langle41\rangle^2}\,,
\qquad
\gamma_L := \frac{\langle43\rangle[\hat13]} {2\langle31\rangle[43]\langle41\rangle^2}\,.
\end{align}
The terms containing $w$ cancel in the complete sum and we obtain
\begin{align}
\chi_L^{abcd} = \frac{1}{2\langle41\rangle^2} u^a u^b u^c u^d\,.
\end{align}

The right core $\chi_R$ simplifies in a similar manner. Writing its four contributions as $\chi_{R,j}$ and with the chosen reference spinors \eqref{eq:LR_ref_spinors}, we find
\begin{align}
\chi_{R,1} &= -\frac{m_n^2}{2\langle41\rangle\langle4\hat2\rangle \langle3\hat2\rangle[34]} \mathcal S[uv^2w]\,, \qquad
&&\chi_{R,2} = -\frac{[24]} {2\langle4\hat2\rangle\langle3\hat2\rangle[34]}\mathcal S[uv^2t]\,, \\
\chi_{R,3} &=\frac{m_n^2\langle34\rangle}
{2\langle41\rangle\langle4\hat2\rangle\langle3\hat2\rangle^2[34]}\mathcal S[v^2tw]\,, \qquad
&& \chi_{R,4}= \frac{\langle34\rangle[24]}
{2\langle4\hat2\rangle\langle3\hat2\rangle^2[34]}\mathcal S[v^2t^2]\,.
\end{align}
We use momentum conservation to write
\begin{align}
-\frac{m_n^2}{\langle41\rangle}w^a -[24]t^a = [34]v^a\,,
\end{align}
Applying this identity to the two pairs of contributions gives
\begin{align}
\chi_{R,1}+\chi_{R,2} = \frac{\mathcal S[v^3u]} {2\langle4\hat2\rangle\langle3\hat2\rangle}\,,\qquad
\chi_{R,3}+\chi_{R,4} = -\frac{\langle34\rangle\,\mathcal S[v^3t]}
{2\langle4\hat2\rangle\langle3\hat2\rangle^2}\,.
\end{align}
Summing all four contributions we get
\begin{align}
\chi_R^{abcd} = \frac{ \mathcal S\!\left[ v^av^bv^c
\left(\langle3\hat2\rangle u^d-\langle34\rangle t^d \right)\right]}
{2\langle4\hat2\rangle\langle3\hat2\rangle^2}\,.
\end{align}
Finally, using Schouten idenity again
\begin{align}
\langle3\hat2\rangle u^a-\langle34\rangle t^a = \langle4\hat2\rangle v^a\,,
\end{align}
we obtain
\begin{align}
\chi_R^{abcd} =\frac{1}{2\langle3\hat2\rangle^2} v^a v^b v^c v^d\,.
\end{align}
In summary, we find
\begin{align}
\chi_L^{abcd} &=\kappa_L\,u^a u^b u^c u^d\,,
&
\chi_{R,abcd} &=\kappa_R\,v_a v_b v_c v_d\,,
\\
\mathbb X_4^{ab} &=c_L\,u^{(a}w^{b)}\,,
&
\mathbb X_{2,ab} &=c_R\,v_{(a}t_{b)},
\end{align}
where
\begin{align}
\kappa_L =\frac{1}{2\langle41\rangle^2}\,, \quad
\kappa_R =\frac{1}{2\langle3\hat2\rangle^2}\,, \quad
c_L =-\frac{m_n}{\langle41\rangle}\,,\quad
c_R =-\frac{m_n}{\langle3\hat2\rangle}\,.
\end{align}
These expressions provide the chiral building blocks for evaluating the normalized contraction $\mathcal F_n$.

\paragraph{Finding $\mathcal{F}_n$ :}
With these factorized structures of $\chi_L$ and $\chi_R$, the two complete symmetric blocks are therefore proportional to
\begin{align}
    \mathcal{S}\!\left(u^{p+4}w^p\right),
    \qquad \mathcal{S}\!\left(v^{p+4}t^p\right)\,,
\end{align}
with $p:=n-1$.
Let us introduce the scalar contractions
\begin{align}
    A:=u^av_a,\qquad
    B:=u^at_a,\qquad
    C:=w^av_a,\qquad
    D:=w^at_a,
    \qquad
    T_0=\kappa_L\kappa_R\,A^4.
\end{align}
If a matching contains $j$ contractions of type $w$--$t$,
the remaining contraction numbers are fixed:
\begin{align}
    N_{wt}=j,\qquad
    N_{wv}=p-j,\qquad
    N_{ut}=p-j,\qquad
    N_{uv}=4+j\,.
\end{align}
Its scalar contraction is consequently
\begin{align}
    A^{4+j}D^j B^{p-j}C^{p-j} =A^4(AD)^j(BC)^{p-j}\,.
\end{align}

To count its normalized multiplicity, fix the arrangement of
the left block. Among its $p$ slots carrying $w$, choose $j$
to meet the right-hand $t$ spinors; the remaining $p-j$
right-hand $t$ spinors must meet slots carrying $u$.
There are therefore
\begin{align}
    \binom pj\binom{p+4}{p-j}
\end{align}
such arrangements, out of
$\binom{2p+4}{p}$ possible placements of the $p$ right-hand
$t$ spinors. Hence 
\begin{align}\label{eq:LR_gluing_finite_sum}
    \mathcal{F}_n = \frac{T_0(c_Lc_R)^p}{\binom{2p+4}{p}}
    \sum_{j=0}^{p} \binom pj\binom{p+4}{p-j} (AD)^j(BC)^{p-j}\,.
\end{align}
The finite sum is a Jacobi polynomial, given as follows
\begin{align}
    P_p^{(0,4)}(x_n) =\sum_{j=0}^{p}\binom pj\binom{p+4}{p-j}
    \left(\frac{x_n+1}{2}\right)^j \left(\frac{x_n-1}{2}\right)^{p-j}\,.
\end{align}
To see how $\mathcal{F}_n$ can be expressed as the Jacobi polynomial, let us define
\begin{align}
    \Delta:=AD - BC\,,\qquad x_n:=\frac{AD + BC}{AD - BC}\,.
\end{align}
Hence
\begin{align}
    \frac{x_n+1}{2}=\frac{AD}{\Delta}\,, \qquad \frac{x_n-1}{2}=\frac{BC}{\Delta}\,.
\end{align}
Therefore, the sum over $j$ in \eqref{eq:LR_gluing_finite_sum} becomes
\begin{align}
    \sum_{j=0}^{p} \binom pj\binom{p+4}{p-j} (AD)^j(BC)^{p-j} = \Delta^p P_p^{(0,4)}(x_n)\,.
\end{align}
and thus
\begin{align}
    \mathcal{F}_n =\frac{T_0}{\binom{2p+4}{p}} (c_Lc_R\Delta)^p P_p^{(0,4)}(x_n)\,.
\end{align}
We express $\Delta , T_0$ and $x_n$ in spinor helicity basis as
\begin{align}
 \Delta =m_n^2\langle 4 1 \rangle\langle3\hat2\rangle\,,\qquad  c_Lc_R\Delta = \frac{m_n^2} {\langle4\hat1\rangle\langle3\hat2\rangle} \Delta = m_n^4\,,\qquad     T_0=\frac{1}{4}[12]^2\langle34\rangle^2\,,
\end{align}
and
\begin{align}
    x_n &:=\frac{ad+bc}{\Delta} = \frac{\langle 43 \rangle\langle 1 \hat2\rangle
    + \langle 4 \hat2 \rangle\langle 1 3 \rangle}{\langle4 1 \rangle\langle 3 \hat2 \rangle} = \left(\frac{2\langle 12 \rangle [12]}{m_n^2} -1\right) \,.
\end{align}
Therefore, after the SU(2) contraction over the internal leading Regge states between the left and right subamplitudes, we obtain
\begin{align}
    \mathcal{F}_n  = [12]^2\langle34\rangle^2 \frac{m_n^{4(n-1)}\Gamma (n) \Gamma (n+4)}{4 \Gamma (2 n+3)} P_{n-1}^{(0,4)}\left(\frac{2\langle 12 \rangle [12]}{m_n^2} -1\right)\,.
\end{align}

Combining massless in \eqref{eq:R0_LR_amp} and massive levels in \eqref{eq:LR_massive_levels}, we get the leading Regge contributions to the BCFW recursion \eqref{eq:LR_gluing_s14_channel} as follows
\begin{align}\label{eq:leading_regge_jacobi_kernel}
A^{\mathrm{LR}}(1^+,2^+,3^-,4^-) =  \alpha'^2  [12]^2\langle34\rangle^2  \left[ \sum_{n=1}^\infty \frac{n^{n-2} (n+2) (n+3)}{2 \Gamma (2 n+2) (\tau + n)} P_{n-1}^{(0,4)}\left(\frac{2 s}{n} -1\right) + \frac{1}{s \tau }\right]\,,
\end{align}
using the definitions of the Mandelstam variables as 
 \begin{align}\label{eq:mandelstams_massless}
     s_{12}:=2k_1\cdot k_2 \equiv \frac{s}{\alpha'} \,,\qquad
     s_{14}= 2k_1\cdot k_4 \equiv \frac{\tau}{\alpha'}\,.
 \end{align}

\subsection{Regge behaviour}\label{sec:LR_Regge}

We study the Regge limit
\begin{align}\label{eq:regge_lr}
    s=-S,\qquad S\to+\infty,\qquad \tau>0\quad\text{fixed}\,,
\end{align}
and separate the helicity factor by writing
\begin{align}
    A^{\mathrm{LR}}(1^+,2^+,3^-,4^-)
    =\alpha'^2[12]^2\langle34\rangle^2 F_{\mathrm{LR}}(-S,\tau)\,,
\end{align}
where
\begin{align}
    & F_{\mathrm{LR}}(-S,\tau) =-\frac{1}{S\tau} +\sum_{n=1}^{\infty}\frac{a_n}{n+\tau}
    P_{n-1}^{(0,4)}\left(-\frac{2S}{n}-1\right)\,,\cr
    &= -\frac{1}{S\tau} - \sum_{n=1}^{\infty}\frac{ (-1)^{n}}{n+\tau} \frac{n^{n-2}(n+2)(n+3)\Gamma(n+4)}{48 \Gamma(2n+2)\Gamma(n)} {}_2F_1\left(1-n,n+4;5;-\frac Sn\right)\,.
\end{align}
We rewrite this as follows
\begin{align}\label{eq:FLR_int1}
    F_{\mathrm{LR}}(-S,\tau) =-\frac{1}{S\tau}-\sum_{n=1}^{\infty}(-1)^n f(n;S,\tau),
\end{align}
where 
\begin{align}
    f(z;S,\tau) &:=\frac{z^{z-2}(z+2)(z+3)\Gamma(z+4)}{48(z+\tau)\Gamma(2z+2)\Gamma(z)}{}_2F_1\left(1-z,z+4;5;-\frac Sz\right),
\end{align}
with principal branches for complex $z$.

\paragraph{Contour deformation:} 
We use a Lindel\"of-type contour representation
\cite{Flajolet2009LindelfRA}, with analytic continuation of Jacobi polynomial
to complex-degree using by its hypergeometric representation
\cite{Cohl2023MultiintegralRF}. This retains the exact level
dependence and permits an analysis by contour deformation.
To do this, first we note that
\begin{align}
    \operatorname*{Res}_{z=n}\frac{\pi}{\sin(\pi z)} =(-1)^n\,,
\end{align}
and define
\begin{align}\label{eq:LR_contour_integrand}
    \mathcal I(z;S,\tau) &:=\frac{\pi}{\sin(\pi z)}f(z;S,\tau)\cr
    &=\frac{z^{z-2}(z+2)(z+3)\Gamma(z+4)\Gamma(1-z)}{48(z+\tau)\Gamma(2z+2)}
    {}_2F_1\left(1-z,z+4;5;-\frac Sz\right)\,,
\end{align}
where we used Euler's reflection formula
\begin{align}
  \Gamma(z)\Gamma(1-z) =  \frac{\pi}{\sin(\pi z)}\,.
\end{align}

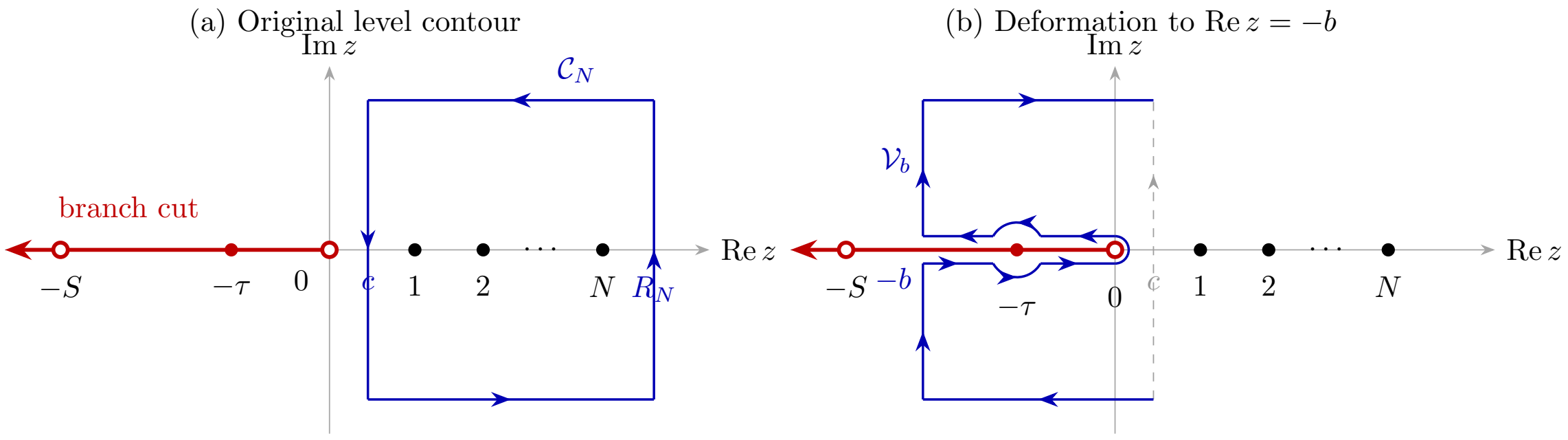
\begin{figure}[t]
\centering
\resizebox{\textwidth}{!}{%
\begin{tikzpicture}[
    >=Stealth,
    font=\small,
    contour/.style={
        thick,blue!70!black,
        postaction={decorate},
        decoration={
            markings,
            mark=at position 0.5 with {\arrow{Stealth}}
        }
    },
    cut/.style={red!75!black,line width=1.3pt},
    reference/.style={
        gray!75,dashed,
        postaction={decorate},
        decoration={
            markings,
            mark=at position 0.75 with {\arrow{Stealth}}
        }
    }
]

\begin{scope}

\node at (0.3,2.65)
    {(a) Original level contour};

\draw[->,gray!70] (-3.8,0) -- (4.45,0)
    node[right,black] {$\operatorname{Re}z$};
\draw[->,gray!70] (0,-2.15) -- (0,2.15)
    node[above,black] {$\operatorname{Im}z$};

\draw[cut,{Stealth}-] (-3.8,0) -- (0,0);
\node[red!75!black,above] at (-2.35,0.25)
    {branch cut};

\draw[cut,fill=white] (-3.15,0) circle (2.4pt);
\draw[cut,fill=white] (0,0) circle (2.4pt);
\fill[red!75!black] (-1.15,0) circle (2.2pt);

\node[below=5pt] at (-3.15,0) {$-S$};
\node[below=5pt] at (-1.15,0) {$-\tau$};
\node[below left=3pt] at (0,0) {$0$};

\foreach \x/\lab in {1/1,1.8/2,3.2/N}{
    \fill (\x,0) circle (2.2pt);
    \node[below=5pt] at (\x,0) {$\lab$};
}
\node at (2.5,0) {$\cdots$};

\draw[contour] (0.45,1.75) -- (0.45,-1.75);
\draw[contour] (0.45,-1.75) -- (3.8,-1.75);
\draw[contour] (3.8,-1.75) -- (3.8,1.75);
\draw[contour] (3.8,1.75) -- (0.45,1.75);

\node[blue!70!black,above] at (2.9,1.82)
    {$\mathcal C_N$};
\node[blue!70!black,below=5pt] at (0.45,0) {$c$};
\node[blue!70!black,below=5pt] at (3.8,0) {$R_N$};

\end{scope}

\begin{scope}[xshift=9.2cm]

\node at (0.3,2.65)
    {(b) Deformation to $\operatorname{Re}z=-b$};

\draw[->,gray!70] (-3.8,0) -- (4.45,0)
    node[right,black] {$\operatorname{Re}z$};
\draw[->,gray!70] (0,-2.15) -- (0,2.15)
    node[above,black] {$\operatorname{Im}z$};

\draw[cut,{Stealth}-] (-3.8,0) -- (0,0);

\draw[reference] (0.45,-1.75) -- (0.45,1.75);
\node[gray!80,below=5pt] at (0.45,0) {$c$};

\foreach \x/\lab in {1/1,1.8/2,3.2/N}{
    \fill (\x,0) circle (2.2pt);
    \node[below=5pt] at (\x,0) {$\lab$};
}
\node at (2.5,0) {$\cdots$};

\draw[contour] (0.45,-1.75) -- (-2.25,-1.75);
\draw[contour] (-2.25,-1.75) -- (-2.25,-0.16);

\draw[contour]
    (-2.25,-0.16) -- (-1.427128,-0.16);
\draw[contour]
    (-1.427128,-0.16)
    arc[start angle=210,end angle=330,radius=0.32];
\draw[contour]
    (-0.872872,-0.16) -- (0,-0.16);

\draw[thick,blue!70!black]
    (0,-0.16)
    arc[start angle=-90,end angle=90,radius=0.16];

\draw[contour]
    (0,0.16) -- (-0.872872,0.16);
\draw[contour]
    (-0.872872,0.16)
    arc[start angle=30,end angle=150,radius=0.32];
\draw[contour]
    (-1.427128,0.16) -- (-2.25,0.16);

\draw[contour] (-2.25,0.16) -- (-2.25,1.75);
\draw[contour] (-2.25,1.75) -- (0.45,1.75);

\draw[cut,fill=white] (-3.15,0) circle (2.4pt);
\draw[cut,fill=white] (0,0) circle (2.4pt);
\fill[red!75!black] (-1.15,0) circle (2.2pt);

\node[below=5pt] at (-3.15,0) {$-S$};
\node[below] at (-1.15,-0.42) {$-\tau$};
\node[below] at (0,-0.30) {$0$};
\node[blue!70!black,left] at (-2.25,-0.35) {$-b$};

\node[blue!70!black,left] at (-2.25,1.05)
    {$\mathcal V_b$};

\end{scope}

\end{tikzpicture}%
}
\caption{
Contour representation of the leading Regge level sum. (a) The counterclockwise contour $\mathcal C_N$ encloses $z=1,\ldots,N$, for some $0<c<1$ and $N<R_N<N+1$.
(b) The upward vertical contour at $\operatorname{Re}z=c$ (dashed)
is deformed to $\operatorname{Re}z=-b$, where $\tau<b<1<S$,
so that it passes to the left of the propagator singularity
at $z=-\tau$ while remaining to the right of the branch point
at $z=-S$.
Open circles denote branch points; filled circles denote poles.
}
\label{fig:LR_contour_deformation}
\end{figure}

We choose the contour $\mathcal{C}_N$ as in fig. \ref{fig:LR_contour_deformation} by noting the following analytic structure of $\mathcal I(z;S,\tau)$ for fixed $S,\tau>0$:
\begin{itemize} 
    \item  we write $z^{z-2}:=\exp\big[(z-2)\operatorname{Log}z\big]$ for $-\pi<\arg z<\pi$ and we choose the principal branch of the hypergeometric function in the segment $[-S,0)$. Together with the cut of $\operatorname{Log}z$, we therefore choose the combined cut along $(-\infty,0]$.
    \item Away from this cut, the only singularities are the simple poles of $\Gamma(1-z)$ at $z=n$, for $n=1,2,\ldots,$ with residues
    \begin{align}
    \operatorname*{Res}_{z=n}\mathcal I(z;S,\tau)
    =(-1)^n f(n;S,\tau)\,.
     \end{align}
   \item The propagator $1/(z+\tau)$ also introduces a singularity at $z=-\tau$, which lies on the chosen cut.
\end{itemize}
For a counterclockwise contour $\mathcal C_N$ enclosing
$z=1,\ldots,N$ and no other singularities, the residue theorem gives
\begin{align}
     \frac{1}{2\pi i}\oint_{\mathcal C_N} dz\,\mathcal I(z;S,\tau) = \sum_{n=1}^{N}(-1)^n f(n;S,\tau)\,.
\end{align}


\paragraph{Contour representation of $F_{\mathrm{LR}}(-S,\tau)$: }
Let the rectangular contour $\mathcal C_N$ have vertices $c\pm iT$ and $R_N\pm iT$,
where $0<c<1$, $R_N=N+1/2$ and $T>0$, enclosing the first $N$ massive levels. Recovering the full level sum requires $N\to\infty$, and hence $R_N\to\infty$. Additionally, we choose $T\to\infty$, and show in appendix \ref{app:exact_contour_reps_FLR} that both contributions decays in these limits and obtain a single vertical-integral representation.

Therefore, the full mass level sum has the exact representation
\begin{align}\label{eq:LR_vertical_contour}
    F_{\mathrm{LR}}(-S,\tau)  = -\frac{1}{S\tau} +\frac{1}{2\pi i} \int_{c-i\infty}^{c+i\infty} dz\,\mathcal I(z;S,\tau)\,, \qquad 0<c<1\,.
\end{align}
Having performed the full level sum through this exact contour representation, we now study its large-$S$ behaviour. We deform the vertical contour from $\operatorname{Re}z=c$ to $\operatorname{Re}z=-b$, choosing $0<\tau<b<1$ and $S>b$. The deformation encounters the branch point at $z=0$ and the propagator singularity at $z=-\tau$, which lies
on the branch cut along negative-real-axis. We therefore retain both boundary values along the cut and indent the contour around these two points as shown in the right panel of Fig. \ref{fig:LR_contour_deformation}.

\paragraph{Large-$S$ behaviour: }
We now use the exact contour representation
\eqref{eq:LR_vertical_contour} to study the limit
$S\to\infty$ at fixed $\tau>0$. 
We use Pfaff's transformation \footnote{For principal branches, Pfaff's transformation holds
for $w\in\mathbb C\setminus[1,\infty)$, with
$|\arg(1-w)|<\pi$; see
\href{https://dlmf.nist.gov/15.8.E1}{NIST DLMF, Eq.~(15.8.1)}.
Under $w=-S/z$, with $S>0$, the excluded set becomes
$[-S,0]$ in the $z$-plane. Together with the principal-branch cut of $z^{z-2}$,
this lies on the chosen common cut $(-\infty,0]$.} to write
\begin{align}
    \mathcal I(z;S,\tau) &= \frac{(z+2)(z+3)\Gamma(z+4)\Gamma(1-z)}
    {48z(z+\tau)\Gamma(2z+2)}(S+z)^{z-1} {}_2F_1\left(1-z,1-z;5;\frac{S}{S+z}\right)\,.
\end{align}
Since ${}_2F_1(1,1;5;1)=4/3$, the integrand behaves near the origin $z=0$ as
\begin{align}
    \mathcal I(z;S,\tau) =\frac{1}{S\tau z}+\mathcal O(z^0)\,.
\end{align}
The corresponding small-circle contribution therefore cancels the massless term in \eqref{eq:FLR_int1} when the contour is deformed to the left.

To describe the remaining cut contribution, let us define
\begin{align}
    \mathcal{N}(z;S)&:=(z+\tau)\mathcal{I}(z;S,\tau),\\
    \operatorname{Disc}_{t}\mathcal{N}(-t;S) &:=\mathcal{N}(-t+i0;S)-\mathcal{N}(-t-i0;S) =2\pi i\,\mathcal{D}(t,S)\,.
\end{align}
For $0<t<\min\{1,S\}$, the hypergeometric discontinuity
formula gives
\begin{align}\label{eq:disc_hypergeometric}
    \mathcal{D}(t,S) &=\frac{(2-t)(3-t)\Gamma(4-t)t^{2-2t}}
    {2\Gamma(2-2t)\Gamma(1+t)\Gamma(4-2t)}(S-t)^{t-4}
    {}_2F_1\left(4-t,4-t;4-2t;-\frac{t}{S-t}\right).
\end{align}
We derive this formula in appendix \ref{app:disc_hypergeometric_2F1}

Now we write down all the integrals after deforming the contour as indicated in the right panel in Fig. \ref{fig:LR_contour_deformation}.
Let us choose $0<\tau<b<1$ and $S>b$. Deforming the contour to
$\operatorname{Re}z=-b$, with symmetric indentations
around $z=-\tau$, gives
\begin{align}\label{eq:LR_contour_decomposition}
    F_{\mathrm{LR}}(-S,\tau) &= \overline{\mathcal N}(-\tau;S)
    +\operatorname{PV}\int_0^b dt\,\frac{\mathcal D(t,S)}{t-\tau}
    +\mathcal V_b(S,\tau)\,,
\end{align}
where
\begin{align}
    \overline{\mathcal N}(-\tau;S) &:= \frac{\mathcal N(-\tau+i0;S)
    +\mathcal N(-\tau-i0;S)}{2}\,,\label{eq:F_LR_parts_1}\\
    \mathcal V_b(S,\tau) &:=\frac{1}{2\pi i}\left(\int_{-b-i\infty}^{-b-i0}
    +\int_{-b+i0}^{-b+i\infty} \right) dz\, \mathcal I(z;S,\tau)\,.\label{eq:F_LR_parts_2}
\end{align}

The first two terms can be estimated directly at large $S$ and we obtain
\begin{align}
    \overline{\mathcal N}(-\tau;S)&=-\frac{\Gamma(\tau)}{\Gamma(1-\tau)}
    S^{-1-\tau}\left[1+\mathcal O(S^{-1})\right],\\
    \operatorname{PV}\int_0^b dt\,\frac{\mathcal D(t,S)}{t-\tau}
    &=\mathcal{O}\left(S^{b-4}\log S\right)\,.
\end{align}
We approximate $\mathcal V_b(S,\tau)$ using the saddle point method in appendix \ref{app:LR_Vu_saddle} and we find
 \begin{align}\label{eq:LR_oscillatory_remainder}
    \mathcal V_b(S,\tau) &= \frac{2}{\sqrt{e\pi}}\, S^{-7/4} \cos\left(2\sqrt{S} + \frac{\pi}{4}\right) + \mathcal{O}\!\left(S^{-b-1}\right)\,,
    \qquad \frac34<b<1\,.
\end{align}
Assembling all the contributions, we find the resulting leading behaviour 
\begin{align}
    F_{\mathrm{LR}}(-S,\tau) \sim -\frac{\Gamma(\tau)}{\Gamma(1-\tau)}
    S^{-1-\tau}\,, \qquad 0<\tau<\frac{3}{4}\,.
\end{align}
We focus on small but fixed positive momentum transfer, $0<\tau<3/4$. This lies within the Regge regime, since $S/\tau\to\infty$ as $S\to\infty$. Within this band, the Regge term scales as $S^{-1-\tau}$, whereas the oscillatory term scales as $S^{-7/4}$ and thus subleading throughout this range.

Including the helicity factor $\alpha'^2 [12]^2 \langle 34\rangle^2 \sim S^2$ in the limit \eqref{eq:regge_lr}, the power $S^{1-\tau}$ identifies the leading Regge trajectory as $\alpha_{\mathrm{LR}}(t)=1+\alpha't$, where $t=-\tau/\alpha'$ is the dimensionful momentum-transfer variable
\begin{align}
     A^{\mathrm{LR}}(1^+,2^+,3^-,4^-)\Big|_{\mathrm{Regge}} \sim  S^{\alpha_{\mathrm{LR}}(t)}\,.
\end{align}

\section{Four-gluon residue at subleading Regge}\label{sec:Four gluon residue at subleading Regge}

We now derive the contribution of the selected subleading
Regge states to the residue of $\mathcal{A}_4(1^+,2^+,3^-,4^-)$. Following the same approach as in the leading Regge case, we reduce the three-point amplitudes derived in section \ref{sec:Three-point amplitudes at subleading Regge} to four dimensions and express them in spinor helicity variables. We then glue these amplitudes using BCFW recursion, with the internal little group contractions yielding residue in terms of Legendre polynomials. Finally, we examine the Regge limit of the corresponding mass-level sum.

\subsection{Dimensional reduction to 4d}

We reduce the ten-dimensional amplitude in \eqref{eq:three-point-subregge} to four dimensions by restricting all momenta and massless gluon polarizations to a four-dimensional
subspace as we did in the previous section:
\begin{align}\label{eq:mom_pol_4d_redux}
    k^A = (k^\mu,\vec 0)\,, \qquad \varepsilon_{\vec k}^{A} = (\varepsilon_{\vec k}^{\mu},\vec 0)\,,
\end{align}
where $A=[0,\ldots,9]$- labels 10d indices, $(\mu ,\nu ,\ldots)=[0,\ldots,3]$- labels 4d indices and internal indices are labeled by $I,J=4,\ldots,9$ that are contracted with $\delta_{IJ}$.
Importantly, this restriction \textit{does not} require the internal components of the massive polarization tensor to vanish.
We parametrize the 4d sector of the massive higher spin polarization in the vertex operator \eqref{eq:massive_pol_subregge} using the symmetric tensor that we used in the leading Regge case \eqref{eq:massive_polarization_spin_n} as follows
\begin{align}\label{eq:massive_pol_subregge}
    \zeta_{\vec k,\mathbf c}^{\rho_1\cdots\rho_n;\nu\rho\sigma}
    :=E_{\vec k,\mathbf c}^{\rho_1\cdots\rho_n} \Omega_{\vec k}^{\nu\rho\sigma},
\end{align}
where $\Omega^{\nu\rho\sigma}_{\vec{k}}$ is a rank-3 antisymmetric tensor.
To satisfy the 10d mixed trace constraints \eqref{eq:mixed trace}, we complete the mixed 4d-6d sector by introducing
\begin{align}\label{eq:SLR_internal_completion}
    \zeta_{\vec k,\mathbf c}^{I\rho_2\cdots\rho_n;J\rho\sigma}
    := -\frac{\delta^{IJ}}{6}\, E_{\vec k,\mathbf c,\lambda}{}^{\rho_2\cdots\rho_n}\Omega_{\vec k}^{\lambda\rho\sigma},
    \qquad n\geq1\,.
\end{align}
Components related by complete symmetry in the first block and antisymmetry in the second block are understood; all other independent components are set to zero.

The BRST transversality in \eqref{eq:BRST_subRegge_NS} imposes
\begin{align}
   k_\nu \Omega^{\nu \rho \sigma}_{\vec{k}} =0\,.
\end{align}
We determine $\Omega^{\nu \rho \sigma}_{\vec{k}}$ as follows. In the rest frame $k^\mu = (m,\vec{0})$, the above condition sets
\begin{align}
    \Omega^{0 ij}_{\vec{k}} =0\,, \qquad \{i,j\}\in [1,2,3]\,.
\end{align}
Thus only the spatial components $\Omega^{ij \ell}_{\vec{k}}$ are non-vanishing.
In three spatial dimensions, a totally anti-symmetric rank-3 little group invariant tensor is proportional to the Levi-Civita tensor: $\Omega^{ijk}_{\vec{k}} \propto \epsilon^{ijk}$. 
Thus, requiring transversality with $k_\mu$, we define
\begin{align}\label{eq:Omega_def}
     \Omega^{\nu \rho \sigma}_{\vec{k}} := \frac{1}{m} k_\mu \epsilon^{\mu \nu \rho \sigma}\,.
\end{align}
Thus, the 4d sector of the massive polarization is
\begin{align}\label{eq:massive_polarization_slr_4d}
    \zeta_{\vec k,\mathbf c}^{\rho_1\cdots\rho_n;\nu\rho\sigma}
    = \frac{1}{m} E_{\vec k,\mathbf c}^{\rho_1\cdots\rho_n} k_\mu \epsilon^{\mu \nu \rho \sigma}\,.
\end{align}
This definition of $\Omega_{\vec{k}}^{\nu \rho \sigma}$ also makes it clear why it carries no SU(2) little-group indices. Indeed, in the rest frame the spatial volume form is invariant under proper rotations
\begin{align}
    R^i{}_{i'} R^j{}_{j'} R^\ell{}_{\ell '} \epsilon^{i'j' \ell'} =
    (\det R)\epsilon^{ij\ell} = \epsilon^{ij\ell}, \qquad R\in SO(3).
\end{align}
Since SU(2) is the double cover of SO(3), $\Omega$ transforms as a SU(2) little group scalar. Thus spin-$n$ dependence of the completed polarization therefore resides entirely in the tensor $E_{\vec k,\mathbf c}$.

The little group for massive particles is defined as the subgroup of Lorentz transformations preserving the standard rest-frame momentum
$\bar k^\mu=(m,\vec0)$. This subgroup is SO(3), whose double
cover is SU(2). Since the rest-frame spatial volume form is
invariant under these rotations, $\Omega$ furnishes a trivial representation under the little group. Boosting it to momentum $k$ gives
$\Omega^{\nu\rho\sigma}_{\vec k} =\frac{1}{m} k_\mu\epsilon^{\mu\nu\rho\sigma}$, without introducing
independent spin labels.

\paragraph{Mixed trace constraint: } We now verify the mixed trace condition explicitly. The full ten-dimensional contraction gives
\begin{align}
    \eta_{AB} \zeta_{\vec k,\mathbf c}^{A\rho_2\cdots\rho_n;B\rho\sigma}
    ={}& \eta_{\mu\nu} \zeta_{\vec k,\mathbf c}^{\mu\rho_2\cdots\rho_n;
    \nu\rho\sigma} + \delta_{IJ}
    \zeta_{\vec k,\mathbf c}^{I\rho_2\cdots\rho_n;J\rho\sigma} \,, \cr
    ={}& E_{\vec k,\mathbf c,\lambda}{}^{\rho_2\cdots\rho_n}
    \Omega_{\vec k}^{\lambda\rho\sigma} -
    \frac{\delta_{IJ}\delta^{IJ}}{6}\,
    E_{\vec k,\mathbf c,\lambda}{}^{\rho_2\cdots\rho_n}
    \Omega_{\vec k}^{\lambda\rho\sigma}\,,\cr
    ={}&0\,,
\end{align}
where we use \eqref{eq:SLR_internal_completion} and $\delta_{IJ}\delta^{IJ}=6$. The mixed 4d-6d sector in the trace does not contribute since the 10d metric is block diagonal: $\eta_{\mu I}=0=\eta_{\nu J}$.
Thus the mixed trace over internal components cancel the nonzero mixed trace of the four-dimensional sector.
For $n\geq2$, another potentially nonzero combination in the mixed trace is the following 
\begin{align}
    \eta_{\alpha\beta} \zeta_{\vec k,\mathbf c}^{\alpha I\rho_3\cdots\rho_n;
    \beta J\sigma} &= \frac{\delta^{IJ}}{6}\, E_{\vec k,\mathbf c,\lambda\beta}{}^{\rho_3\cdots\rho_n} \Omega_{\vec k}^{\lambda\beta\sigma} =0\,.
\end{align}
All remaining mixed traces vanish by the component assignments. The full 10d massive higher spin polarization $\zeta$ therefore satisfies the mixed trace constraint in \eqref{eq:mixed trace}.

\paragraph{Young-symmetry condition : }
Next, we check the BRST constraint \eqref{eq:fermionic transversality_yong}. Let us consider the purely 4d components first
\begin{align}
\zeta_{\vec k,\mathbf c}^{\rho_2\cdots\rho_n;
[\rho_1\nu\rho\sigma]} &= E_{\vec k,\mathbf c}^{[\rho_1|\rho_2\cdots\rho_n|}
\Omega_{\vec k}^{\nu\rho\sigma]}
\cr 
&=- \frac{k_\lambda}{m}\, E_{\vec k,\mathbf c}^{\rho_2\cdots\rho_n[\rho_1} \epsilon^{\nu\rho\sigma]\lambda}\,,
\end{align}
where vertical bars exclude indices within it from antisymmetrization.
From the 4d Schouten identity 
\begin{align}
\epsilon^{[\lambda\nu\rho\sigma} E_{\vec k,\mathbf c}^{\rho_1]\rho_2\cdots\rho_n}=0\,,
\end{align}
after contracting with $k_\lambda$ and using the transversality condition \eqref{eq:E_tensor_transverse_traceless}, we obtain
\begin{align}
k_\lambda E_{\vec k,\mathbf c}^{\rho_2\cdots\rho_n [\rho_1} \epsilon^{\nu\rho\sigma]\lambda}=0\,.
\end{align}
Therefore we have
\begin{align}
\zeta_{\vec k,\mathbf c}^{\rho_2\cdots\rho_n; [\rho_1\nu\rho\sigma]}=0\,,
\end{align}
as required by the last BRST condition in \eqref{eq:fermionic transversality_yong}.
We now check the components involving mixed 4d-6d indices. Suppressing the labels $(\vec k,\mathbf c)$, the Young condition \eqref{eq:fermionic transversality_yong} in 10d can be expressed as follows 
\begin{align}\label{eq:schematic_young}
\mathcal Y^{ABCD}_{\mathbf R} :={}& \zeta^{A\mathbf R;BCD}
-\zeta^{B\mathbf R;ACD}+\zeta^{C\mathbf R;ABD} -\zeta^{D\mathbf R;ABC} =0\,,
\end{align}
where $\mathbf R$ denotes the remaining symmetric Lorentz indices. For the mixed 4d-6d sector \eqref{eq:SLR_internal_completion}, there are two potentially nontrivial cases.
First, when both internal indices ($IJ$) lie among the four antisymmetrized indices, we obtain
\begin{align}
\mathcal Y^{IJ\mu\nu}_{\mathbf r} = \zeta^{I\mathbf r;J\mu\nu}
-\zeta^{J\mathbf r;I\mu\nu}
= -\frac{\delta^{IJ}-\delta^{JI}}{6}
E_{\lambda}{}^{\mathbf r}\Omega^{\lambda\mu\nu}
=0\,,
\end{align}
where $\mathbf r$ contains $n-1$ four-dimensional indices. Due to component assignment of $\zeta$, the last two terms in \eqref{eq:schematic_young} are zero in this case.
Second, when one 6d index lies in the symmetric spectator block, 
\begin{align}
\mathcal Y^{J\mu\nu\rho}_{I\mathbf r} &=
-\zeta^{\mu I\mathbf r;J\nu\rho}
+\zeta^{\nu I\mathbf r;J\mu\rho}
-\zeta^{\rho I\mathbf r;J\mu\nu}\cr
&= \frac{\delta^{IJ}}{6}
\left(E_{\lambda}{}^{\mu\mathbf r}\Omega^{\lambda\nu\rho}
+E_{\lambda}{}^{\nu\mathbf r}\Omega^{\lambda\rho\mu}
+E_{\lambda}{}^{\rho\mathbf r}\Omega^{\lambda\mu\nu}\right)\,,
\end{align}
where we set $A=J$ and $\mathbf{R} = \mathbf r \cup (I,\mu)$ with $\mathbf r$ now contains $(n-2)$ 4d indices and for this component assignment, the first term in \eqref{eq:schematic_young} is zero.
Since $E$ is transverse, the expression in parenthesis can be evaluated in a three-dimensional subspace orthogonal to $k$. Then we can write
\begin{align}
E_{\lambda}{}^{\mu\mathbf r}\Omega^{\lambda\nu\rho}
+
E_{\lambda}{}^{\nu\mathbf r}\Omega^{\lambda\rho\mu}
+
E_{\lambda}{}^{\rho\mathbf r}\Omega^{\lambda\mu\nu}
=
\left(\Pi_{\alpha\beta}E^{\alpha\beta\mathbf r}\right)
\Omega^{\mu\nu\rho} =0\,,
\end{align}
where we define the projector $\Pi_{\alpha\beta}:=\left(\eta_{\alpha\beta} + \frac{k_\alpha k_\beta}{m^2}\right)$ with $k^2=-m^2$, and the last equality follows from transversality and tracelessness of $E$.

All other internal-index assignments involve vanishing components
or are related to these cases by the index symmetries. Thus the full ten-dimensional completion satisfies the BRST conditions in \eqref{eq:fermionic transversality_yong}.

\paragraph{Three-point amplitude in 4d :}

We now extract the three-point amplitude in 4d from the full ten-dimensional expression in \eqref{eq:three-point-subregge} using the representation of the massive polarization in \eqref{eq:massive_polarization_slr_4d} below\footnote{Due to the choice of external momentum and polarization in \eqref{eq:mom_pol_4d_redux} the mixed sector of the massive polarization does not contribute
\begin{align}
     \zeta_{\vec k,\mathbf c}^{I\rho_2\cdots\rho_n;J\rho\sigma} k_{2I}k_{2\rho_2}\cdots k_{2\rho_n} \varepsilon_{2,J}k_{1[\sigma}\varepsilon_{1,\rho]}
    = -\frac{k_2^J}{6}\, E_{\vec k,\mathbf c,\lambda}{}^{\rho_2\cdots\rho_n}\Omega_{\vec k}^{\lambda\rho\sigma}\varepsilon_{2,J}k_{1[\sigma}\varepsilon_{1,\rho]}=0\,.
\end{align}
Note that, the full 10d transversality of gluon polarization implies $k_{2J}\varepsilon_{2}^{J}\neq 0$.
}
\begin{align}\label{eq:SLR_three-point_4d}
    \mathcal{A}_3(1^1_{\mathrm{LR}}, 2^{1}_{\mathrm{LR}}, \mathbf{3}_{\mathrm{SLR}}^n) &= (-1)^n (2\alpha')^{n/2}  (\zeta_{3,\nu\rho \sigma}\cdot k_2^n) \left(\varepsilon_2^\nu k_1^{[\sigma}\varepsilon_1^{\rho]} - \varepsilon_2^\rho k_1^{[\sigma}\varepsilon_1^{\nu]} + \varepsilon_2^\sigma k_1^{[\rho}\varepsilon_1^{\nu]} \right)\,,\cr
    &=  6(-1)^n \frac{(2\alpha')^{n/2}}{m_3 n!} (\varepsilon_3 \cdot k_2)^{\odot n} \epsilon(k_1,\varepsilon_1,k_2,\varepsilon_2) \,,
\end{align}
with $\sum_{i=1}^3 k_i =0$ and $\epsilon(a,b,c,d):=\epsilon_{\mu \nu \rho \sigma} a^\mu b^\nu c^\rho d^\sigma$. We made the SU(2) little group indices carried by the spin-1 massive polarizations $\varepsilon^\mu_{\vec k,c_1c_2}$ implicit in the definition
\begin{align}
    (\varepsilon_3 \cdot k_2)^{\odot n} := \mathcal{S}_{c_1 \cdots c_{2n}} \left[ \varepsilon^{(\rho_1}_{3,c_1c_2} \cdots \varepsilon^{\rho_n)}_{3,c_{2n-1}c_{2n}} \right] k_2^{\rho_1} \cdots k_2^{\rho_n}\,.
\end{align}
We restore the SU(2) indices when we rewrite the three-point amplitude in  spinor helicity formalism in the next section.

We now show that, the opposite helicity sector of this three-point amplitude vanishes identically.
Let us rewrite $\epsilon(k_1,\varepsilon_1,k_2,\varepsilon_2)$ as follows
\begin{align}
    \epsilon(k_1,\varepsilon_1,k_2,\varepsilon_2) &= \frac{1}{4} \epsilon_{\mu \nu \rho \sigma} k_1^{[\mu} \varepsilon_1^{\nu]} k_2^{[\rho}\varepsilon_2^{\sigma]}= -\frac{1}{4}\epsilon_{\mu \nu \rho \sigma} F_1^{\mu \nu} F_{2}^{\rho \sigma}\,.
\end{align}
where the field strength in momentum space is $F_{j,\mu \nu}= i k_{j,[\mu}\varepsilon^{(j)}_{\nu]}$ for $j-$th particle. Going to the spinor helicity basis for massless particles \footnote{This can be seen as follows
\begin{align}
    F_{\mu \nu}=\bar{\sigma}_\mu^{\dot{\alpha}\alpha}\bar{\sigma}_\nu^{\dot{\beta}\beta}F_{\alpha \dot{\alpha}\beta \dot{\beta}}=i k_{[\mu}\varepsilon_{\nu]} = i \bar{\sigma}_{[\mu}^{\dot{\alpha}\alpha}\bar{\sigma}_{\nu]}^{\dot{\beta}}\lambda_\alpha \tilde{\lambda}_{\dot{\alpha}}\varepsilon_{\beta \dot{\beta}}
    \Rightarrow F_{\alpha \dot{\alpha}\beta \dot{\beta}} =i(\lambda_\alpha \tilde{\lambda}_{\dot{\alpha}}\varepsilon^h_{\beta \dot{\beta}} - \lambda_\beta \tilde{\lambda}_{\dot{\beta}}\varepsilon^h_{\alpha \dot{\alpha}})\,.
\end{align}
Using $\varepsilon^+_{\alpha \dot{\alpha}}(k)=\frac{q_\alpha \tilde{\lambda}_{\dot{\alpha}}}{\sqrt{2}\langle q\lambda \rangle}$ and $\varepsilon^-_{\alpha \dot{\alpha}}(k)=\frac{\lambda_{\alpha}\tilde{q}_{\dot{\alpha}}}{\sqrt{2}[ q\lambda ]}$, we get
\begin{align}
     F^+_{\alpha \dot{\alpha}\beta \dot{\beta}} &= \frac{i}{\sqrt{2}}\tilde{\lambda}_{\dot{\alpha}}\tilde{\lambda}_{\dot{\beta}} \epsilon_{\alpha\beta}=: \frac{i}{\sqrt{2}} \tilde{F}^{+}_{\dot{\alpha}\dot{\beta}}\epsilon_{\alpha\beta} \,, \cr 
     F^-_{\alpha \dot{\alpha}\beta \dot{\beta}} &= \frac{i}{\sqrt{2}}\lambda_{\alpha}\lambda_{\beta} \epsilon_{\dot{\alpha}\dot{\beta}} =:\frac{i}{\sqrt{2}} F^{-}_{\alpha \beta} \epsilon_{\dot{\alpha}\dot{\beta}}\,.
\end{align}
Note that, for opposite helicity choice: $F^+ =0\,, \tilde{F}^-=0$.
Combining both helicity contributions, we obtain the identity below.
}
\begin{align}
    F_i^{\mu \nu} &= \bar{\sigma}^{\dot{\alpha}\alpha,[\mu} \bar{\sigma}^{\nu],\dot{\beta}\beta} \left( \lambda_{i\alpha} \lambda_{i\beta} \epsilon_{\dot{\alpha} \dot{\beta}} +  \tilde{\lambda}_{i\dot{\alpha}} \tilde{\lambda}_{i\dot{\beta}} \epsilon_{\alpha \beta} \right)\,,
\end{align}    
we find
\begin{align}
    \epsilon(k_1,\varepsilon^{h_1}_1,k_2,\varepsilon^{h_2}_2) 
    = -\kappa \left( F^{h_1}_{1\alpha \beta} F_2^{h_2,\alpha \beta} - \tilde{F}^{h_1}_{1,\dot{\alpha} \dot{\beta}} \tilde{F}_2^{h_2,\dot{\alpha} \dot{\beta}} \right)\,.
\end{align}
using
\begin{align}
   \epsilon_{\mu \nu \rho \sigma}\bar{\sigma}^{\mu\dot{\alpha}\alpha} \bar{\sigma}^{\nu\dot{\beta}\beta} \bar{\sigma}^{\rho\dot{\gamma}\gamma} \bar{\sigma}^{\sigma\dot{\delta}\delta} =\kappa \left[\epsilon^{\alpha \beta}\epsilon^{\dot{\alpha}\dot{\beta}} \epsilon^{\gamma \delta}\epsilon^{\dot{\gamma}\dot{\delta}} - \epsilon^{\alpha \gamma}\epsilon^{\dot{\alpha}\dot{\gamma}} \epsilon^{\beta \delta}\epsilon^{\dot{\beta}\dot{\delta}} - \epsilon^{\alpha \delta}\epsilon^{\dot{\alpha}\dot{\delta}} \epsilon^{\gamma \beta}\epsilon^{\dot{\gamma}\dot{\beta}}\right]\,,
\end{align}
with $\kappa = 4i$ for $\epsilon_{0123}=+1$. Therefore, in the opposite helicity sector
\begin{align}
    \epsilon(k_1,\varepsilon^+_1,k_2,\varepsilon^-_2) = -\kappa \left( F^+_{1\alpha \beta} F_2^{-,\alpha \beta} - \tilde{F}^+_{1\dot{\alpha} \dot{\beta}} \tilde{F}_2^{-,\dot{\alpha} \dot{\beta}} \right)=0\,,
\end{align}
and hence the three-point amplitude vanishes identically
\begin{align}
 \mathcal{A}(1^+, 2^{-}, \mathbf{3}^n)=0\,.    
\end{align}
In the same helicity sector, we find
\begin{align}
    \epsilon(k_1,\varepsilon^+_1,k_2,\varepsilon^+_2) &= \kappa  \tilde{F}^+_{1\dot{\alpha} \dot{\beta}} \tilde{F}_2^{+,\dot{\alpha} \dot{\beta}} = -2i [12]^2\,,\cr
    \epsilon(k_1,\varepsilon^-_1,k_2,\varepsilon^-_2) &=- \kappa  F^{-}_{1{\alpha} {\beta}} F_2^{-,{\alpha} {\beta}}= -2i \langle 12\rangle^2\,.
\end{align}
Therefore, the three-point amplitudes in the same helicity sectors are given by
\begin{align}\label{eq:slr_amp3_same_helicity}
    \mathcal{A}_3(1^+_{\mathrm{LR}}, 2^{+}_{\mathrm{LR}}, \mathbf{3}_{\mathrm{SLR}}^n) &=(-1)^{n+1} \frac{12 i}{m_3 n!} (2\alpha')^{n/2} (\varepsilon_3 \cdot k_2)^{\odot n} [12]^2 \,,\cr
     \mathcal{A}_3(1^-_{\mathrm{LR}}, 2^{-}_{\mathrm{LR}}, \mathbf{3}_{\mathrm{SLR}}^n) &= (-1)^{n+1}\frac{12 i}{m_3 n!}  (2\alpha')^{n/2}(\varepsilon_3 \cdot k_2)^{\odot n} \langle 12\rangle^2 \,.
\end{align}

\subsection{BCFW shift and four-point residue}

We now consider the contribution to the colour-ordered amplitude $\mathcal A_4(1^+,2^+,3^-,4^-)$ at subleading Regge.
The $[1^+\,2^+\rangle$ shift used in the leading Regge calculation
uses the $s_{14}$ channel in which gluons of opposite helicity are coupled
to each cubic vertex. Since these couplings vanish for the
subleading Regge states, we instead use the $[3^-\,2^+\rangle$ shift. This uses the $s_{12}$ channel, where the two cubic amplitudes involve the nonvanishing $(1^+,2^+)$ and $(3^-,4^-)$ gluon couplings.
To apply this shift, we complexify the following external momenta
\begin{align}
\hat k_3^\mu=k_3^\mu-z \, r^\mu\,, \qquad \hat k_2^\mu=k_2^\mu+z\,r^\mu\,,
\end{align}
and realized in terms of spinor helicity variables as
\begin{align}
|\hat3]&=|3]-z|2]\,,
&
|\hat2\rangle&=|2\rangle+z|3\rangle\,,
\\
|\hat3\rangle&=|3\rangle\,,
&
|\hat2]&=|2]\,,
\end{align}
with $r_{\alpha\dot\alpha}
=|3\rangle_\alpha[2|_{\dot\alpha}$. In this channel, the intermediate momentum is $\hat P_n(z) = k_1+\hat k_2(z) = -(\hat k_3(z)+k_4)$. All shifted spinor helicity brackets are evaluated at $z=z_{12}$. To determine it
at mass level $n$, we set the complex propagator on shell
\begin{align}
 (\hat{s}_{12} + m_n^2)_{z=z_{12}}=0\, \Rightarrow z_{12,n} = -\frac{(2k_1\cdot k_2 + m_n^2)}{2r\cdot k_1}\,, \qquad m_n^2=\frac{n+1}{\alpha'}.
\end{align}

We denote the contribution of the selected subleading Regge
states to the residue by
\begin{align}\label{eq:slr_gluing_s12_channel}
\mathcal R_n^{\mathrm{SLR}}(1^+,2^+,3^-,4^-) :={}& \left.\mathcal A_3 (1^+_{\mathrm{LR}},\hat2^+_{\mathrm{LR}},
-\hat{\mathbf P}_{n,\mathrm{SLR}})
\mathcal A_3 (\hat3^-_{\mathrm{LR}},4^-_{\mathrm{LR}},\hat{\mathbf P}_{n,\mathrm{SLR}}) \right|_{z=z_{12,n}},
\end{align}
where contraction of the internal SU(2) little-group indices
implements the intermediate-state sum. The corresponding contribution to the BCFW recursion is\footnote{There is no massless exchange at subleading Regge. The $n=0$ state is a massive pseudoscalar.}
\begin{align}
A^{\mathrm{SLR}}(1^+,2^+,3^-,4^-) =\sum_{n=0}^{\infty} \frac{\mathcal R_n^{\mathrm{SLR}}(1^+,2^+,3^-,4^-)}{s_{12}+m_n^2}\,.
\end{align}
This selects the subleading Regge states from the intermediate-state
sum of the full string amplitude.


Substituting the spinor helicity representation of the massive spin-1 polarization, we express the deformed three-point amplitudes \eqref{eq:slr_amp3_same_helicity} as follows
\begin{align}\label{eq:slr_amp3_same_helicity_spinor}
\mathcal{A}_3(1^+_{\mathrm{LR}},\hat{2}_{\mathrm{LR}}^+,-\hat{\mathbf{P}}_{n,\mathrm{SLR}}) &=  (-1)^{n+1}\frac{12 i}{m_P n!}(2\alpha')^{n/2} (-\varepsilon_{-\vec{P}} \cdot k_1)^{\odot n} [12]^2  \,,\cr
&=  (-1)^{n+1}\frac{12 i}{m_P n!} (2\alpha')^{n/2} \left( \frac{i}{\sqrt{2}m_P }\right)^{n} \langle \hat{\mathbf{P}}|k_1|\hat{\mathbf{P}}]^{\odot 2n} [12]^2\,,\cr
\mathcal{A}_3(\hat{3}^-_{\mathrm{LR}},4^-_{\mathrm{LR}},\hat{\mathbf{P}}_{n,\mathrm{SLR}}) &=  (-1)^{n+1}\frac{12 i}{m_P n!} (2\alpha')^{n/2} (\varepsilon_{\vec{P}} \cdot k_4)^{\odot n} \langle 34\rangle^2 \,,\cr
 &=  (-1)^{n+1}\frac{12 i}{m_P n!} (2\alpha')^{n/2} \left( \frac{i}{\sqrt{2}m_P}\right)^{n} \langle \hat{\mathbf{P}}|k_4|\hat{\mathbf{P}}]^{\odot 2n} \langle 34\rangle^2 \,.
\end{align}
The SU(2) indices left implicit in the bold face notation $\langle \hat{\mathbf{P}}|k_1|\hat{\mathbf{P}}]^{\odot 2n}$, defined as follows
\begin{align}\label{eq:symmetric_spin_state_polarization}
   \langle \hat{\mathbf{P}}|k_1|\hat{\mathbf{P}}]^{\odot 2n}&= \frac{1}{\mathcal{N}_n} \sum_{(i,j)\in \mathcal{S}'_n} \langle \hat{P}^{(a_{i_1}}|k_1|\hat{P}^{a_{j_1})}] \cdots \langle \hat{P}^{(a_{i_n}}|k_1|\hat{P}^{a_{j_n})}]\,,
\end{align}
where $\mathcal{S}'_n$ includes only the distinct pairings and the normalization is given by
\begin{align}
    \mathcal{N}_n:= (2n-1)!! = \frac{(2n)!}{2^n n!}\,.
\end{align}

Next, we consider the little group contraction of these totally symmetric rank-$2n$ tensors. Before we give the formula for general $n$,
let us evaluate the SU(2) contraction with some low values of $n$. For $n=1$:
\begin{align}\label{eq:slr_gluing_n=1_int}
      \langle \hat{\mathbf{P}}|k_1|\hat{\mathbf{P}}]^{\odot 2}  \langle \hat{\mathbf{P}}|k_4|\hat{\mathbf{P}}]_{\odot 2} &= - 8( m_P^2 (k_1\cdot k_4) + (k_1\cdot \hat{P})(k_4\cdot \hat{P}))\,,\cr
  &= - 8 m_P^2\left(  k_1\cdot k_4 - \frac{m_P^2}{4} \right)\,,
\end{align}
in the $s_{12}$ channel with $\hat{P}= k_1+k_2 = -(k_3+k_4)$. Here, we evaluate shifted scalar products at $z=z_{12}$ as $k_1\cdot \hat{k}_2 = -m_P^2/2= k_4\cdot \hat{k}_3$ and $\alpha' m_P^2 = (n+1)$.
For $n=2$, we find
\begin{align}\label{eq:slr_n23_gluing}
     \langle \hat{\mathbf{P}}|k_1|\hat{\mathbf{P}}]^{\odot 4}  \langle \hat{\mathbf{P}}|k_4|\hat{\mathbf{P}}]_{\odot 4} &=  64 m_P^4 (k_1\cdot k_4)^2 + 128 m_P^2 (k_1\cdot k_4) (k_1\cdot \hat{P})(k_4\cdot \hat{P}) \cr
     & \hspace{5cm} +\dfrac{128}{3} (k_1\cdot \hat{P})^2(k_4\cdot \hat{P})^2\cr
     &=64 m_P^4 (k_1\cdot k_4)^2 - 32 (k_1\cdot k_4) m_P^6 +\frac{8}{3}m_P^8 \,.
\end{align}

\paragraph{General $n$ case: }

In order to organize the SU(2) little group contractions, let us consider again the $n=2$ case. We define the following symmetric rank-2 tensors
\begin{align}
    \quad X^{ab}:= \langle \hat{P}^{(a}|k_1|\hat{P}^{b)}] \,, \qquad \quad Y_{ab}:= \langle \hat{P}_{(a}|k_4|\hat{P}_{b)}]\,.
\end{align}
We consider these tensors as basic building blocks to express the massive spinor brackets, for example
\begin{align}
    \langle \hat{\mathbf{P}}|k_1|\hat{\mathbf{P}}]^{\odot 4}  \equiv  \mathbb{X}^{abcd}\,,\qquad
    \langle \hat{\mathbf{P}}|k_4|\hat{\mathbf{P}}]_{\odot 4}  \equiv \mathbb{Y}_{abcd}\,,
\end{align}
where we define
\begin{align}
    \mathbb{X}^{abcd} &= \frac{1}{3}\left( X^{ab}X^{cd} + X^{ac}X^{bd}+ X^{ad}X^{bc} \right)\,,\cr
     \mathbb{Y}_{abcd} &= \frac{1}{3}\left( Y_{ab}Y_{cd} + Y_{ac}Y_{bd}+ Y_{ad}Y_{bc} \right)\,, 
\end{align}
Making use of the symmetry property of $X,Y$, the little group contraction gives
\begin{align}
    \mathbb{X}^{abcd} \mathbb{Y}_{abcd} = \frac{1}{9} \left( 3 X^{ab}X^{cd}  Y_{ab}Y_{cd} + 6  X^{ab}X^{cd} Y_{ac}Y_{bd} \right)\,.
\end{align}
The first term evaluates to
\begin{align}
    X^{ab}X^{cd}  Y_{ab}Y_{cd} = -[8\{ m_n^2 (k_1\cdot k_4) + (k_1\cdot \hat{P})(k_4\cdot \hat{P})\}]^2= (X\cdot Y)^2\,.
\end{align}
Note that the bilinear $X^{ab}Y_{ab}$ corresponds to the $n=1$ case $\langle \hat{\mathbf{P}}|k_1|\hat{\mathbf{P}}]^{\odot 2}  \langle \hat{\mathbf{P}}|k_4|\hat{\mathbf{P}}]_{\odot 2}$ defined in \eqref{eq:slr_gluing_n=1_int} and is equal to the scalar product $(X\cdot Y)$ in the above expression.
To evaluate the second term in $\mathbb{X}^{abcd} \mathbb{Y}_{abcd}$, we first rearrange this as follows. Consider 
\begin{align}
    X^{ab}X^{cd}-X^{ac}X^{bd} = \xi \epsilon^{ad}\epsilon^{bc}\,.
\end{align}
LHS is anti-symmetric in $(a,d)$ and $(b,c)$ and therefore should be proportional to $\epsilon^{ad}\epsilon^{bc}$ in two dimensions. To determine $\xi$, we contract both sides with $\epsilon_{ad}\epsilon_{bc}$
 \begin{align}
     (X^{ab}X^{cd}-X^{ac}X^{bd})\epsilon_{ad}\epsilon_{bc} = 2 X^{ab} X_{ab} = 4\xi \Rightarrow \xi =\frac{1}{2}X^{ab} X_{ab} = \frac{X^2}{2}\,.  
 \end{align}
Therefore we obtain the identity satisfied by the rank-2 tensors
\begin{align}\label{eq:product_X2_Scouten_rule}
     X^{ab}X^{cd}-X^{ac}X^{bd} =  \frac{X^2}{2} \epsilon^{ad}\epsilon^{bc}\,.
\end{align}
The SU(2) index contraction of LHS with $Y_{ac}Y_{bd}$ gives 
\begin{align}
    X^{ab}X^{cd} Y_{ac}Y_{bd} = (X\cdot Y)^2 - \frac{1}{2}X^2 Y^2 \,,
\end{align}
where $Y^2:= Y^{ab}Y_{ab}$ and the LHS is the same as the second term of $\mathbb{X}^{abcd} \mathbb{Y}_{abcd}$ . 
Collecting these results, we find
\begin{align}
    \mathbb{X}^{abcd} \mathbb{Y}_{abcd} &= \frac{1}{9} \left( 3 X^{ab}X^{cd}  Y_{ab}Y_{cd} + 6  X^{ab}X^{cd} Y_{ac}Y_{bd} \right)\,,\cr
    &= (X\cdot Y)^2 - \frac{1}{3}X^2 Y^2\,.
\end{align}
Introducing mass-level dependent scalar products with $\alpha' m_P^2 = \alpha' m_n^2 = (n+1)$ at subleading Regge trajectory
\begin{align}\label{eq:slr_su2_invaraints}
    u_n &:= X\cdot Y = -8[m_n^2 (k_1\cdot k_4) + (k_1\cdot \hat{P})(k_4\cdot \hat{P})]= 2m_n^4 \left(1  -\frac{4 k_1\cdot k_4}{m_n^2}\right)\,,\cr
    x_n&:= X^2 =-8 (k_1\cdot \hat{P})^2 =-8(k_1\cdot \hat{k}_2)^2 = -2m_n^4\,,\cr
    y_n&:=Y^2 =-8 (k_4\cdot \hat{P})^2=-8 (k_4 \cdot \hat{k}_3)^2 = -2m_n^4\,,
\end{align}
we write the $n=2$ contribution as follows
\begin{align}
     \mathbb{X}^{abcd} \mathbb{Y}_{abcd} &= u_2^2 - \frac{1}{3} x_2y_2\,,\cr
     &= 64 m_P^4 (k_1\cdot k_4)^2 + 128 m_P^2 (k_1\cdot k_4) (k_1\cdot \hat{P})(k_4\cdot \hat{P})
     +\frac{128}{3} (k_1\cdot \hat{P})^2(k_4\cdot \hat{P})^2\cr
     &= \langle \hat{\mathbf{P}}_2|k_1|\hat{\mathbf{P}}_2]^{\odot 4}  \langle \hat{\mathbf{P}}|k_4|\hat{\mathbf{P}}]_{\odot 4}\,.
\end{align}
In the second step, we kept the mass label as $m_P$ to compare with the first step in \eqref{eq:slr_n23_gluing}. Thus we recover the $n=2$ case that we found earlier. 

The explicit $n=1,2$ calculations illustrate how the Schouten identity organizes the massive spinor-bracket contractions into the invariants: $u_n$, $x_n$, and $y_n$ defined in \eqref{eq:slr_su2_invaraints}.These SU(2) invariants give the spinor helicity realization of the spin-$n$ projector in the polarization space discussed in \cite{Chiodaroli:2021eug}. Its Legendre polynomial representation
then yields the general $n$ result
\begin{align}\label{eq:genreal_n_su2_gluing}
    \langle \hat{\mathbf{P}}|k_1|\hat{\mathbf{P}}]^{\odot 2n}  \langle \hat{\mathbf{P}}|k_4|\hat{\mathbf{P}}]_{\odot 2n} &= \frac{n!}{(2n-1)!!} (x_n y_n)^{\frac{n}{2}} \mathrm{P}_n\left(\frac{u_n}{\sqrt{x_n y_n}}\right)\,,\cr
    &= \frac{n!}{(2n-1)!!} (2m_n^4)^{n} \mathrm{P}_n\left( 1- \frac{4 k_1\cdot k_4}{m_n^2} \right)\,.
\end{align}
where $\mathrm{P}_n$ is the Legendre polynomial.
We give an inductive proof of this formula in appendix \ref{app:SLR_gluing}.
This completes the SU(2) little group contractions. 

Next, we use these relations to obtain the contribution to the four-point residue at subleading Regge trajectory \eqref{eq:slr_gluing_s12_channel}
\begin{align}
   \mathcal R_n^{\mathrm{SLR}}(1^+,2^+,3^-,4^-) = -144 [12]^2 \langle 34\rangle^2 \frac{(-2i\alpha')^n}{2 (n!)^2 m_n^{2n+2}}  \langle \hat{\mathbf{P}}|k_1|\hat{\mathbf{P}}]^{\odot 2n}  \langle \hat{\mathbf{P}}_n|k_4|\hat{\mathbf{P}}_n]_{\odot 2n}\,.
\end{align}
We relabel $m_P=m_n$ in three-point amplitudes in \eqref{eq:slr_amp3_same_helicity_spinor} to display the mass-level dependence explicitly. Substituting the general $n$ result \eqref{eq:genreal_n_su2_gluing}, we obtain
\begin{align}\label{eq:subleading_regge_legendre_kernel}
    A^{\mathrm{SLR}}(1^+,2^+,3^-,4^-) = - 144 \alpha'^2 [12]^2 \langle 34\rangle^2 \sum_{n=0}^\infty  \frac{(-4)^n}{(2n)!} \frac{(n+1)^{n-1}}{s+n+1} \mathrm{P}_n\left(1 - \frac{2\tau}{n+1} \right)\,,
\end{align}
using the dimensionless Mandelstam variables defined in \eqref{eq:mandelstams_massless}.

\subsection{Regge behaviour}

We now examine the Regge behavior of the selected SLR contribution to the same color-ordered amplitude $\mathcal{A}_4(1^+,2^+,3^-,4^-)$. The $[1^+\,2^+\rangle$ shift used in the LR calculation accesses the $s_{14}$ factorization channel, whereas the $[3^-\,2^+\rangle$ shift used in the SLR calculation uses the $s_{12}$ channel. To examine the SLR sum in the corresponding exchange-channel Regge limit, we interchange the roles of the energy and momentum-transfer variables relative to the LR analysis: $s_{14}$ serves as the energy invariant, while $s_{12}$ is held fixed as the momentum transfer. In our dimensionless variables, this amounts to taking the limit
\begin{align}\label{eq:regge_slr}
\tau =-\mathcal S,\qquad \mathcal S\to +\infty\,, \qquad s=\sigma>0\quad\text{fixed}\,.
\end{align}

We accordingly analyze $F_{\mathrm{SLR}}(\sigma,-\mathcal{S})$ defined as follows
\begin{align}
    A^{\mathrm{SLR}}(1^+,2^+,3^-,4^-) = -144\alpha'^2[12]^2\langle34\rangle^2 F_{\mathrm{SLR}}(\sigma,-\mathcal{S})\,,
\end{align}
where
\begin{align}
    F_{\mathrm{SLR}}(\sigma,-\mathcal{S}) &= \sum_{n=0}^{\infty} \frac{b_n}{\sigma+n+1} \mathrm{P}_n\left( -\frac{2\mathcal{S}}{n+1}-1 \right)\,,\\
    b_n&:=\frac{4^n}{\Gamma(2n+1)}(n+1)^{n-1}\,.
\end{align}
Below we present a shorter, alternative analysis to the contour method used in the leading Regge case. The leading term in the large $\mathcal{S}$ expansion is
\begin{align}
    b_n  \mathrm{P}_n\left( -\frac{2\mathcal{S}}{n+1}-1 \right) \xrightarrow{\mathcal{S}\to \infty} \frac{(- 4 \mathcal{S})^n}{(n!)^2 (n+1)}\,,
\end{align}
for fixed $n$.
Now, let $x'_{n,j}$ be the Legendre zeros, we then write
\begin{align}
     b_n  \mathrm{P}_n\left( -\frac{2\mathcal{S}}{n+1}-1 \right) = \frac{(-4 \mathcal{S})^n}{(n!)^2 (n+1)} Q'_n(\mathcal{S})\,,
\end{align}
with
\begin{align}
    Q'_n(\mathcal{S}):= \prod_{j=1}^n \left(1- \frac{\rho'_{n,j}}{\mathcal{S}} \right)\,, \qquad \rho'_{n,j}:= -\frac{(n+1)}{2}(1+x'_{n,j})\,,
\end{align}
with $Q'_0(\mathcal{S})=1$.
Using $\frac{1}{\sigma +n +1}= \int_0^1 du\, u^{n+\sigma}$, we write
\begin{align}
    F_{\mathrm{SLR}}(\sigma,-\mathcal{S}) &= \int_0^1 du\, u^{\sigma} \sum_{n=0}^\infty \frac{(-4 \mathcal{S} u)^n}{(n!)^2 (n+1)} Q'_n(\mathcal{S})
\end{align}
With a change of integration variable $y= 4 \mathcal{S} u$ and replacing $Q'_n(\mathcal S)$ by its fixed-level limit defines
the leading-power approximation
\begin{align}
    F_{\mathrm{SLR}}^{\mathrm{app}}(\sigma,-\mathcal S)
    &:=
    \frac{1}{(4\mathcal S)^{\sigma+1}}
    \int_0^{4\mathcal S}dy\,
    y^{\sigma-\frac12}J_1(2\sqrt y)
    \notag\\
    &=
    (4\mathcal S)^{-1-\sigma}
    \frac{\Gamma(1+\sigma)}{\Gamma(1-\sigma)}
    +\mathcal O(\mathcal S^{-5/4}),
    \qquad 0<\sigma<\frac14.
\end{align}

Including the helicity factor $\alpha'^2 |[12]^2 \langle 34\rangle^2| = \sigma^2$ which is fixed in the limit \eqref{eq:regge_slr}, we find
\begin{align}
     A^{\mathrm{SLR}}(1^+,2^+,3^-,4^-)\Big|_{\mathrm{Regge}} \sim  \mathcal{S}^{-1+\alpha' t}\,, \qquad 0< \sigma < 1/4\,.
\end{align}
with $t=-\frac{\sigma}{\alpha'}$. This leading-power approximation exhibits the Regge exponent associated with this selected SLR states.

A contour analysis of the exact level sum confirms the leading-power result. The leading oscillatory correction is of order $\mathcal S^{-5/4}$, and is therefore subleading to the Regge contribution for $0<\sigma<1/4$.

 To compare  the LR and SLR trajectories, both the masses and the four-dimensional spins must be taken into account:
\begin{align}
\begin{aligned}
\mathrm{LR}:&\qquad \alpha'm_n^2=n\,,
&J_n&=n+1\,,\\
\mathrm{SLR}:&\qquad \alpha'm_n^2=n+1\,,
&J_n&=n\,.
\end{aligned}
\end{align}
Eliminating $n$, we obtain $J_{\mathrm{LR}}=\alpha'm^2+1$ and $J_{\mathrm{SLR}}=\alpha'm^2-1$, corresponding to
\begin{align}
\alpha_{\mathrm{LR}}(t)=1+\alpha't\,, \qquad \alpha_{\mathrm{SLR}}(t)=-1+\alpha't\,.
\end{align}
Thus the intercepts differ by two: at a fixed mass level, the selected SLR states lie two spin units below the leading trajectory.

\section{Conclusion}\label{sec:conclusion}

We have constructed the contributions of selected leading and subleading Regge states to the four-gluon amplitude by gluing their three-point amplitudes through the BCFW recursion. The resulting Jacobi and Legendre kernels make the dependence on the exchanged states explicit and reveal a differential relation between the two families.

After removing the common factor $\alpha'^2[12]^2\langle34\rangle^2$ and the channel propagators, let us define
\begin{align}
    \widehat{\mathcal K}^{\mathrm{LR}}_n(x) &=a_nP_{n-1}^{(0,4)}(x)\,,&
    \widehat{\mathcal K}^{\mathrm{SLR}}_{n+1}(x) &=b_{n+1}P_n(x)\,,
\end{align}
where the subscripts denote the mass levels. The Jacobi degree and parameter relations yield a differential operator $\mathcal O_n$ satisfying
\begin{align}
    \mathcal O_nP_{n-1}^{(0,4)}(x)&=P_n(x)\,,&
    \widehat{\mathcal K}^{\mathrm{SLR}}_{n+1}(x) &=\frac{b_{n+1}}{a_n}\,
    \mathcal O_n\widehat{\mathcal K}^{\mathrm{LR}}_n(x)\,,
    \qquad n\geq1 \,.
\end{align}
The channel-dependent expressions follow by substituting $x=2s/n-1$ for LR and $x=2\tau/(n+1)-1$ for SLR and restoring their respective propagators. We give the details in appendix \ref{app:kernel_relations}.

This relation suggests a possible way to organize the remaining intermediate-state contributions. Our broader aim is to reconstruct the full string amplitude solely using three-point data and on-shell recursion, using relations
among the kernels to organize the intermediate-state sum. An important next step is to investigate whether analogous maps extend to the other Regge families and reproduce, with the appropriate couplings and state multiplicities,
the known four-point worldsheet amplitude.

In a forthcoming work, we will extend this construction to amplitudes with massive external states, using covariant recursion relations based on massive--massless shifts \cite{Falkowski:2020aso, Ballav:2020ese,Ballav:2021ahg}.

\section*{Acknowledgement}

During the preparation of this manuscript, we used OpenAI's
ChatGPT to improve the exposition and assist with algebraic checks. All scientific ideas, methods,
derivations, and conclusions were developed and verified
by the author and assumes full responsibility for the accuracy and scientific content.

\begin{appendix}

\section{OPEs in RNS superstring}\label{app:OPE_RNS_superstrings}

We use the conventions of \cite{Schlotterer:2011psa}. We list the standard matter holomorphic part of the OPE's below. We start with the basic OPEs between the bosonic and fermionic matter fields
\begin{align}\label{OPE_matter_fields}
     X^\mu(z)X^\nu (w) &\sim -2\alpha'\eta^{\mu \nu} \ln|z-w| \,,\qquad && \psi^\mu(z)\psi^\nu (w) \sim \frac{\eta^{\mu \nu}}{z-w} \,,\cr    \partial X^\mu (z)X^\nu (w)  & \sim -\frac{2 \alpha' \eta^{\mu \nu}}{(z-w)}\,, \qquad  && \partial \psi^\mu(z)\psi^\nu (w) \sim -\frac{\eta^{\mu \nu}}{(z-w)^2} \,,\cr
    \partial X^\mu (z)  \partial X^\nu (w) & \sim -\frac{2 \alpha' \eta^{\mu \nu}}{(z-w)^2}\,,\qquad && \psi^\mu(z) \partial \psi^\nu (w) \sim \frac{\eta^{\mu \nu}}{(z-w)^2}\,.
\end{align}
The matter stress tensor in the RNS formulation is 
\begin{align}\label{eq:stress_tensor_matter_rns}
    T_m(z) =-\frac{1}{4\alpha'}:\partial X^\mu (z)\partial X_\mu(z):+\frac{1}{2}:\partial \psi^\mu(z)\psi_\mu(z):\,.
\end{align}
The OPE between this stress tensor and various matter fields are 
\begin{align}
    T_m(z)\partial X^\mu(w) &\sim \frac{\partial_w X^\mu(w)}{(z-w)^2}+\frac{\partial^2_wX^\mu(w)}{z-w}\,,\cr
    T_m(z) \psi^\mu(w) &\sim \frac{(1/2) \psi^\mu(w)}{(z-w)^2}+\frac{\partial \psi^\mu(w)}{z-w}\,,\cr
    T_m(z)\partial \psi^\mu(w) & \sim \frac{\psi^\mu(w)}{(z-w)^3}+\frac{3}{2}\frac{\partial \psi^\mu(w)}{(z-w)^2}+\frac{\partial^2 \psi^\mu (w)}{z-w}\,.
\end{align}
All the operators in RHS are normal ordered.
The matter supercurrent in the RNS formulation of superstring theory is 
\begin{align}\label{eq:supercurrent_def_rns}
    G(z)=\frac{1}{4\sqrt{2\alpha'}}:i\partial X_\mu (z)\psi^\mu (z):\,.
\end{align}
Its OPEs with the matter fields are given below
\begin{align}\label{eq:supercurrent_opes}
    G(z)i\partial X^\mu (0) &\sim \sqrt{\frac{\alpha'}{2}} \left[\frac{(1/2) \psi^\mu (0)}{z^2}+\frac{(1/2)\partial \psi^\mu (0)}{z} \right]\cr
    G(z)i\partial^2 X^\mu (0) &\sim \sqrt{\frac{\alpha'}{2}} \left[\frac{\partial \psi^\mu (0)}{z^2}+\frac{\psi^\mu(0)}{z^3}+\frac{(1/2)\partial^2 \psi^\mu (0)}{z} \right]\cr
    G(z)\psi^\mu (0) &\sim \frac{1}{4\sqrt{2\alpha'}} \frac{i \partial X^\mu (0)}{z} \cr
    G(z) \partial \psi^\mu (0) &\sim \frac{1}{4\sqrt{2\alpha'}}\left[\frac{i \partial^2 X^\mu(0)}{z}+\frac{i \partial X^\mu (0)}{z^2}\right]\,.
\end{align}
Using the bosonic OPE
\begin{align}
   i \partial X^\mu (z)e^{ik\cdot X(w)}\sim 2\alpha' k^\mu \frac{e^{ik\cdot X(w)}}{z-w} \,.
\end{align}
we get
\begin{align}\label{sup_curr_planwave_ope}
    G(z)e^{ik\cdot X(w)} &\sim \sqrt{\frac{\alpha'}{2}} \frac{k_\mu \psi^\mu (w)}{2}\frac{e^{ik\cdot X(w)}}{z-w} \,.
\end{align}

\section{Pomeron vertex derivation}\label{app:Pomeron_vertex}

\subsection{Pomeron in tachyon scattering}

We recall the derivation of the Pomeron vertex operator in tachyon scattering amplitude in bosonic string theory.
Consider the OPE of two tachyon vertex operators
\begin{align}
	\hat{\mathcal{A}}(z)_{\text{t}-\text{t}} =\int_{0}^1 dw \, e^{i\hat{k}_1\cdot X(0)}\, e^{i\hat{k}_2\cdot X(w)}\,.
\end{align}    
We use
\begin{align}\label{eq:pomeron_plane_wave}
    e^{i \hk_1 \cdot X(0)} e^{i \hk_2 \cdot X(w)} &= w^{2\alpha' k_1 \cdot k_2} e^{ik \cdot X(0) + iw \hat{k}_2\cdot \dot{X}(0)} \left[ 1+ \frac{i w^2}{2}\hat{k}_2 \cdot \ddot{X}(0) + \mathcal{O}(w^3)\right]\,,
\end{align}
to rewrite $\hat{\mathcal{A}}(z)_{\text{t}-\text{t}}$ as follows
\begin{align}\label{ope_prod_1}
		 \hat{\mathcal{A}}(z)_{\text{t}-\text{t}} &= \int_0^1 d w \, w^{\alpha' s_{12}- 2} e^{ik \cdot X(0)+iw \hat{k}_2\cdot \dot{X}(0)} \left[ 1+ \frac{i w^2}{2}\hat{k}_2 \cdot \ddot{X}(0) + \mathcal{O}(w^3)\right]\,.
\end{align}
Here we have used $\hat{k}_1\cdot \hat{k}_2 =k_1\cdot k_2$ and define $s_{12}:=(k_1+k_2)^2 = (2k_1\cdot k_2 + 2/\alpha')$ and $k:=(k_1+k_2)$. The key insight of BPST \cite{Brower:2006ea} was that the $w$ integral dominates near the region $w\sim \frac{1}{z}$. Therefore, the term $w\hat{k}_2\cdot \dot{X}(0)$ is $\mathcal{O}(1)$ and all other terms in the OPE are $\mathcal{O}\left(\frac{1}{z}\right)$ and can therefore be neglected at large $z$. Performing the $w$ integral in this regime, we obtain
\begin{align} \label{pomeron_vert}
	\hat{\mathcal{A}}(z)_{\text{t}-\text{t}} &= \int_0^{1} d w \, w^{\alpha' s_{12}-2} e^{i(k_1 +k_2)\cdot X(0)+iw \hat{k}_2\cdot \dot{X}(0)}\cr
        &= \left(-i\hat{k}_2\cdot \dot{X}(0)\right)^{1- \alpha' s_{12}} \Gamma \left[ \alpha' s_{12}-1\right]e^{ik \cdot X(0)} \,,
\end{align}
for $(1- \alpha' s_{12}) < 0$. This hybrid operator $\mathcal{P}$ is called the Pomeron vertex operator \cite{Brower:2006ea}
\begin{align}
    \mathcal{P}_{\mathrm{t}} :=(-i\hat{k}_2\cdot \dot{X}(0))^{1 - \alpha' s_{12}} \Gamma \left[\alpha' s_{12}-1\right]e^{ik \cdot X(0)} \,, \quad \text{for}\quad  \mathrm{Re}(1- \alpha' s_{12})<0\,.
\end{align}
In the case of superstrings we need to include the superghost factor $e^{-2\phi(0)}$ into the above expression to obtain \eqref{eq:poemron_vertex_massless}.

\subsection{Massless gauge boson}

Here we derive \eqref{eq:pomeron_VV_ope}. Recall that the gluon vertex operator in open superstring in the canonical $(-1)$ ghost picture is
\begin{align}
	V_i^{(-1)}(k_i,z_i)=\varepsilon_{i,A} \psi^A(z_i) e^{i k_i \cdot X(z_i)}e^{-\phi(z_i)} \,.
\end{align}
Consider
\begin{align}
    V_1^{(-1)}(\hat{k}_1,0)\, V_2^{(-1)}(\hat{k}_2,w) \sim \varepsilon_{1,A}\varepsilon_{2,B} \left[ \psi^A(0) \psi^B (w)\right] \left[ e^{i \hk_1 \cdot X(0)} e^{i \hk_2 \cdot X(w)} \right] \left[ e^{-\phi (0)} e^{-\phi (w)} \right]\,.
\end{align}
For the superghost sector, we use
\begin{align}
    e^{-\phi(0)}e^{-\phi(w)} &= -\frac{1}{w} :e^{-2\phi(0)}\left[1-w\dot{\phi}(0)+\frac{w^2}{2}\big((\dot\phi)^2-\ddot\phi\big) \right]: + \mathcal{O}(w^2)\,,
\end{align}
and for the matter sectors we use \eqref{eq:pomeron_plane_wave} and 
\begin{align}
    \psi^A(0)\psi^B (w) & = -\frac{\eta^{A B}}{w} + :\psi^A \psi^B: (0) + w\, :\psi^A \partial \psi^B: (0) +\mathcal{O}(w^2)\,,
\end{align}
to write
\begin{align}
    (\hat{\varepsilon}_1 \cdot \psi(0)) (\hat{\varepsilon}_2 \cdot \psi(w)) &= - \frac{\hat{\varepsilon}_1 \cdot \hat{\varepsilon}_2}{w} + :\left[(\hat{\varepsilon}_1 \cdot \psi ) (\hat{\varepsilon}_2 \cdot \psi) + w(\hat{\varepsilon}_1\cdot \psi) (\hat{\varepsilon}_2 \cdot \dot{\psi}) \right]:(0) + \mathcal{O}(w^2)\,.
\end{align}
We obtain
\begin{align}
      V_1^{(-1)}(\hat{k}_1,0)\, V_2^{(-1)}(\hat{k}_2,w) =  w^{\alpha' s_{12}-2} e^{ik \cdot X(0)+iw \hat{k}_2\cdot \dot{X}(0)} e^{-2\phi(0)}\left(1-w\dot{\phi}(0)\right)\cr
      \times \left[ (\hat{\varepsilon}_1 \cdot \hat{\varepsilon}_2) -w :(\hat{\varepsilon}_1 \cdot \psi ) (\hat{\varepsilon}_1 \cdot \psi):(0) + \mathcal{O}(w^2) \right]\cr
      \times \left[ 1+ \frac{i w^2}{2}\hat{k}_2 \cdot \ddot{X}(0) + \mathcal{O}(w^3)\right]\,.
\end{align}
Rearranging terms, we can rewrite this expression as follows
\begin{align}\label{eq:VV_pomeron_OPE}
     V_1^{(-1)}(\hat{k}_1,0)\, & V_2^{(-1)}(\hat{k}_2,w) =  w^{\alpha' s_{12}-2} e^{ik \cdot X+iw \hat{k}_2\cdot \dot{X}} e^{-2\phi} \cr
     & \times \Big[  (\hat{\varepsilon}_1 \cdot \hat{\varepsilon}_2) \left(1 -w \dot{\phi} + \frac{i w^2}{2}\hat{k}_2 \cdot \ddot{X} \right) -w (\hat{\varepsilon}_1 \cdot \psi ) (\hat{\varepsilon}_2 \cdot \psi) \left(1-w \dot{\phi}\right)\cr
     & \hspace{2cm} - w^2 (\hat{\varepsilon}_1\cdot \psi) (\hat{\varepsilon}_2 \cdot \dot{\psi}) +\frac{w^2}{2}(\hat{\varepsilon}_1 \cdot \hat{\varepsilon}_2) (\dot{\phi}^2 - \ddot{\phi}) + \mathcal{O}(w^3) \Big]\,.
\end{align}
As mentioned earlier, the Pomeron integral is dominated near $w\sim z^{-1}$ in the large-$z$ regime. Since, the deformed momentum $\hat{k}_2 =\mathcal{O}(z)$, the above expansion must be organized simultaneously in $w$ and the shifted momentum both. This can be seen as follows. Consider the following integral 
\begin{align}
    \int_0^1 dw\, w^{\alpha' s_{12}-2+r} e^{iw \hat{k}_2\cdot \dot{X}} \sim (-i \hat{k}_2\cdot \dot{X})^{1-\alpha' s_{12}-r}\,.
\end{align}
and thus after integrating over $w$, the omitted terms in \eqref{eq:VV_pomeron_OPE} are suppressed by at least $\mathcal{O}(z^{-2})$ relative to the leading exponent $(-i\hat{k}_2\cdot\dot{X})^{1-\alpha's_{12}}$, with the shifted polarizations stripped off.

\section{Vanishing of same helicity sector at LR in 4d}\label{app:same_helicity_LR}

The three-point amplitude in leading Regge with same helicity gluons vanishes \cite{Feng:2011qc}. This can be seen as follows. Recall that the rank-4 core tensor in the leading Regge three-point amplitude is defined in\eqref{eq:core_tensor_3pt_LR}
\begin{align}
    \chi_{(h_1,h_2)} = \Bigg[ \frac{n}{2\alpha'} (\varepsilon_2^{h_2} \cdot \varepsilon_{3}) (\varepsilon_1^{h_1} \cdot \varepsilon_{3}) + \left\{ -(\varepsilon_2^{h_2} \cdot k_1) (\varepsilon_1^{h_1} \cdot \varepsilon_{3})+ (\varepsilon_2^{h_2} \cdot \varepsilon_{3}) (\varepsilon_1^{h_1} \cdot k_2) \right\}(\varepsilon_{3}\cdot k_1)\cr
    + (\varepsilon_2^{h_2} \cdot \varepsilon_1^{h_1} )(\varepsilon_{3}\cdot k_1)^2 \Bigg]\,.
\end{align}
We show this vanishes in the same helicity sector.
Using the polarization vectors for massive spin-1 and massless particles in \eqref{eq:massive_polarizations_spinorhelicity}, \eqref{eq:massless_polarizations_spinorhelicity}, for both positive helicity
\begin{align}
    \chi^{abcd}_{(+,+)} = -\frac{1}{2m_3^2}\frac{1}{\langle r_2 2\rangle \langle r_1 2\rangle} \mathcal{S} \Bigg[\frac{n}{\alpha'} \langle r_2 3^{a}\rangle \langle r_1 3^b\rangle [23^c][13^d] + \langle 3^a 1\rangle [13^b] [21]\big\{ \langle r_12\rangle \langle r_2 3^c\rangle [23^d] \cr
    +\langle r_2 1\rangle \langle r_1 3^c\rangle [13^d]\big\} + \langle 3^a 1\rangle \langle 3^b 1\rangle [13^c][13^d] [21]\langle r_2 r_1\rangle  \Bigg]\,.
\end{align}
Here $\alpha' m_3^2= n$ at leading Regge and $r_{1,2}$ are the reference spinors with positive helicity polarizations. The symmetrization over all little group indices are understood: $\mathcal{S}\equiv \mathcal{S}_{abcd}$.
The second and third terms inside the braces can be simplified as
\begin{align}
    &\langle r_12\rangle \langle r_2 3^c\rangle [23^d] 
    +\langle r_2 1\rangle \langle r_1 3^c\rangle [13^d] = \langle r_1 |2|3^d] \langle r_23^c\rangle + \langle r_2 1\rangle \langle r_1 3^c\rangle [13^d] \cr
    &=\left( -\langle r_1 1\rangle \langle r_23^c\rangle + \langle r_2 1\rangle \langle r_1 3^c\rangle \right) [13^d] - m_3 \langle r_1 3^d\rangle \langle r_2 3^c\rangle \cr
    &= \langle r_1 r_2 \rangle \langle 3^c 1\rangle [13^d] - m_3 \langle r_1 3^d\rangle \langle r_2 3^c\rangle\,,
\end{align}
using Schouten identity. Substituting back in $\chi$, we get
\begin{align}
    \chi^{abcd}_{(+,+)} = -\frac{1}{2m_3^2}\frac{1}{\langle r_2 2\rangle \langle r_1 2\rangle} \mathcal{S} \Bigg[\frac{n}{\alpha'} \langle r_2 3^{a}\rangle \langle r_1 3^b\rangle [23^c][13^d] + \langle 3^a 1\rangle [13^b] [21]\big\{\langle r_1 r_2 \rangle \langle 3^c 1\rangle [13^d]\cr
    - m_3 \langle r_1 3^d\rangle \langle r_2 3^c\rangle\big\} + \langle 3^a 1\rangle \langle 3^b 1\rangle [13^c][13^d] [21]\langle r_2 r_1\rangle  \Bigg]\,.
\end{align}
Since $\chi$ is totally symmetric in the four SU(2) indices, the second and last terms (with a common factor $\langle r_1 r_2\rangle$) cancel 
\begin{align}
    \chi^{abcd}_{(+,+)} = -\frac{1}{2m_3^2}\frac{1}{\langle r_2 2\rangle \langle r_1 2\rangle} \mathcal{S}\Bigg[ \langle r_2 3^{a}\rangle \langle r_1 3^b\rangle \left( \frac{n}{\alpha'}  [23^c][13^d] - m_3 \langle 3^a 1\rangle [13^b] [21] \right)\Bigg]\,.
\end{align}
Using $\langle 3^a 1\rangle [21]= - \langle 3^a |k_1|2] = \langle 3^a |k_3|2]= -m_3 [3^a2]$\,, we get
\begin{align}
    \chi^{abcd}_{(+,+)} = -\frac{1}{2m_3^2}\frac{1}{\langle r_2 2\rangle \langle r_1 2\rangle} \mathcal{S}\Bigg[ \langle r_2 3^{a}\rangle \langle r_1 3^b\rangle[23^c][13^d] \Bigg]\left( \frac{n}{\alpha'}  - m_3^2 \right)=0\,,
\end{align}
using mass-shell condition. Therefore, the 4d amplitude in \eqref{eq:three-point-LR-4d} for both positive helicity gluons is vanishing. Similarly, both negative helicity sector also vanishes.

\section{SU(2) contraction at subleading Regge}\label{app:SLR_gluing}

To establish the general result, we proceed by induction on the
number of rank-2 building blocks. At a fixed mass level $n$, let us define
\begin{align}
G_r^{(n)}:= \mathbb X_{(r)}^{a_1\cdots a_{2r}} \mathbb Y_{(r),a_1\cdots a_{2r}}\,,
\end{align}
where
\begin{align}
\mathbb X_{(r)}^{a_1\cdots a_{2r}} &:=
\mathcal S_{a_1\cdots a_{2r}}
\left[X^{a_1a_2}\cdots X^{a_{2r-1}a_{2r}}\right]\,,
\\
\mathbb Y_{(r),a_1\cdots a_{2r}} &:=\mathcal S_{a_1\cdots a_{2r}}
\left[Y_{a_1a_2}\cdots Y_{a_{2r-1}a_{2r}} \right]\,.
\end{align}
Here $\mathcal S$ denotes symmetrization over the SU(2) indices including the factor $\frac{1}{(2r-1)!!}$, and $r$ labels the tensor rank independently of the fixed mass level $n$.
The spin-projector recurrence in eq. 2.19 of \cite{Chiodaroli:2021eug} gives
\begin{align}\label{eq:su2_gluing_recurrence}
G_{r+1}^{(n)}= u_n G_r^{(n)} -\frac{r^2}{(2r-1)(2r+1)}x_ny_n G_{r-1}^{(n)}\,,
\end{align}
with
\begin{align}
G_0^{(n)}=1,\,, \qquad G_1^{(n)}=u_n\,.
\end{align}
We claim that
\begin{align}\label{eq:su2_gluing_induction}
G_r^{(n)} = \frac{r!}{(2r-1)!!} (x_ny_n)^{r/2} \mathrm P_r\left( \frac{u_n}{\sqrt{x_ny_n}}\right)\,.
\end{align}
The initial cases $r=0,1$ follow immediately, with
$(-1)!!:=1$. Suppose that \eqref{eq:su2_gluing_induction}
holds at ranks $r$ and $r-1$. Writing
$t_n:=u_n/\sqrt{x_ny_n}$, substituting $G_r^{(n)}$ into \eqref{eq:su2_gluing_recurrence} yields
\begin{align}
G_{r+1}^{(n)} &=\frac{r!}{(2r-1)!!}
(x_ny_n)^{(r+1)/2} \left[ t_n\mathrm P_r(t_n) -\frac{r}{2r+1}\mathrm P_{r-1}(t_n)\right]\,.
\end{align}
Using the Legendre recurrence relation
\begin{align}
(2r+1)t\,\mathrm P_r(t) = (r+1)\mathrm P_{r+1}(t) +r\mathrm P_{r-1}(t)\,,
\end{align}
we find
\begin{align}
G_{r+1}^{(n)} &= \frac{(r+1)!}{(2r+1)!!} (x_ny_n)^{(r+1)/2} \mathrm P_{r+1}(t_n)\,.
\end{align}
This completes the induction. Setting $r=n$ in \eqref{eq:su2_gluing_induction} establishes \eqref{eq:genreal_n_su2_gluing}.

\section{Regge limit of LR contribution}\label{app:Regge limit of LR contribution}

\subsection{Large $N$ and $T$ estimates of $\mathcal{I}$}\label{app:exact_contour_reps_FLR}

To estimate the integrand on the horizontal edges, we write $z=x+it = r e^{i\theta}$ with $x\geq c>0$. We keep $S,\tau>0$ fixed and use the following bounds for sufficiently large $r$ and $|t|$:
\begin{itemize}
    \item Using Euler's reflection formula we write: $\Gamma(z+4)\Gamma(1-z) =z(z+1)(z+2)(z+3)\frac{\pi}{\sin(\pi z)}$, so that the prefactor in \eqref{eq:LR_contour_integrand} satisfies
    \begin{align}
        \left| \frac{z(z+1)(z+2)^2(z+3)^2}{48(z+\tau)} \right| = \mathcal{O}(r^5)\,.
    \end{align}

    \item Since $|\sin\pi(x+it)|^2 =\sin^2(\pi x)+\sinh^2(\pi t)$, we have $\left|\frac{\pi}{\sin(\pi z)}\right| = \mathcal{O}(e^{-\pi|t|})$. This can be seen as follows
    \begin{align}
        \left|\frac{\pi}{\sin(\pi z)}\right| = \frac{\pi}{(\sin^2(\pi x)+\sinh^2(\pi t))^{1/2}} \leq \frac{\pi}{\sinh(\pi |t|)} = \frac{2 \pi e^{-\pi |t|}}{1-e^{-2\pi |t|}}\,.
    \end{align}
    Using the principal logarithm $\operatorname{Log}z=\log r+i\theta$, we obtain $|z^{z-2}|=r^{x-2}e^{-t\theta}$.

    \item From the Stirling's approximation we find
    \begin{align}
        \frac{1}{|\Gamma(2z+2)|} \leq (2r)^{-2x-3/2}e^{2x+2t\theta}\,.
    \end{align}

    \item Finally, the hypergeometric function obeys the fixed-$S$ bound\cite{https://doi.org/10.1002/mma.208}
    \begin{align}
        \left| {}_2F_1\left(1-z,z+4;5;-\frac Sz\right) \right| = \mathcal{O}(e^{6 \sqrt{Sr}})\,.
    \end{align}
\end{itemize}


This can be seen by considering the the Gauss series representation
\begin{align}
    {}_2F_1\left(1-z,z+4;5;-\frac Sz\right) &=\sum_{j=0}^{\infty}u_j\,,&
    u_j&:=\frac{(1-z)_j(z+4)_j}{(5)_j\,j!}\left(-\frac Sz\right)^j\,.
\end{align}
For fixed $S>0$, let $r=|z|\geq\max\{5,8S\}$ and $m=\lceil r\rceil$. For $j\leq m$, each Pochhammer factor in numerator has modulus at most $3r$\footnote{We have $|(1-z)_j|\leq (3r)^j$ and $|(z+4)|_j \leq (3r)^j$ for $j\leq m$.}. Since $(5)_j\geq j!$, we obtain
\begin{align}
    \sum_{j=0}^{m}|u_j| \leq\sum_{j=0}^{\infty}\frac{(9Sr)^j}{(j!)^2}
    \leq\left(\sum_{j=0}^{\infty}\frac{(3\sqrt{Sr})^j}{j!}\right)^2
    =e^{6\sqrt{Sr}}\,.
\end{align}
For $j\geq m$, the successive terms satisfy
\begin{align}
    |u_{j+1}|&\leq\frac Sr
    \frac{(r+j+1)(r+j+4)}{(j+5)(j+1)}|u_j|\leq\frac{4S}{r}|u_j|
    \leq\frac12|u_j|\,.
\end{align}
Thus $\sum_{j=m+1}^{\infty}|u_j|\leq |u_m| \leq e^{6\sqrt{Sr}}$, and therefore
\begin{align}
    \left|{}_2F_1\left(1-z,z+4;5;-\frac Sz\right)\right| =  \mathcal{O}(e^{6\sqrt{Sr}})\,.
\end{align}
This bound holds uniformly in $\arg z$ for sufficiently large $r$ at fixed $S$. Combining these estimates gives
\begin{align}\label{eq:LR_contour_decay_bound}
    |\mathcal I(z;S,\tau)| = \mathcal{O}\left(r^{3/2-x} \exp\left[(2-\log4)x+t\theta-\pi|t|+C_S\sqrt r \right]\right)\,,
\end{align}
where $C_S \sim \sqrt{S}$.
For fixed $N,S,\tau$, each horizontal contribution decays at large $T$
\begin{align}
    \left|\int_c^{R_N}dx\,\mathcal I(x\pm iT;S,\tau)\right| =\mathcal{O}\left(
    T^{3/2-c}\exp\left[-\frac{\pi T}{2}+6\sqrt{ST}\right]\right)  \longrightarrow0\,.
\end{align}
Here $r=\sqrt{x^2 +T^2}\sim T$ and $t\theta \leq \frac{\pi T}{2}$.

On the right vertical edge, $x=R_N = N+1/2$. For sufficiently large
$N$, we have
\begin{align}
    r^{3/2-R_N}\leq R_N^{3/2-R_N}\,, \qquad \sqrt r\leq\sqrt{R_N}+\sqrt{|t|}\,,
\end{align}
and therefore the integral also decays at large $N$
\begin{align}
    \int_{-\infty}^{\infty}dt\,
    |\mathcal I(R_N+it;S,\tau)| =\mathcal{O}\left(R_N^{3/2}\exp\left[-R_N\log R_N+(2-\log4)R_N+6\sqrt{SR_N}\right]\right) \longrightarrow 0\,.
\end{align}
 The decay follows because $-R_N\log R_N$ dominates the remaining terms at fixed $S,\tau$.

With this choice of contour choice, taking $T\to\infty$ for fixed $N$ first and then $N\to\infty$ leaves only the left vertical edge
\begin{align}
    \lim_{N\to\infty}\lim_{T\to\infty} \oint_{\mathcal C_N}dz\,\mathcal I(z;S,\tau) = -\int_{c-i\infty}^{c+i\infty} dz\,\mathcal I(z;S,\tau)\,.
\end{align}

\subsection{Derivation of $ \mathcal{D}(t,S) $}\label{app:disc_hypergeometric_2F1}
This can be derived as follows. Let us define
\begin{align}
    \mathcal G(z;S) &:={}_2F_1\left(1-z,1-z;5;\frac{S}{S+z}\right)\,, &q(z)&:=\frac{z}{S+z}\,.
\end{align}
Using the connection formula relating the hypergeometric
functions around arguments $0$ and $1$, we obtain
\begin{align}\label{eq:LR_hypergeom_connection}
    \mathcal G(z;S) = &\frac{24\Gamma(3+2z)}{\Gamma(4+z)^2}
    {}_2F_1(1-z,1-z;-2z-2;q(z))\cr
    & \qquad + \frac{24\Gamma(-3-2z)}{\Gamma(1-z)^2} q(z)^{3+2z}{}_2F_1(4+z,4+z;4+2z;q(z))\,.
\end{align}
For fixed $0<t<b$ and $z=-t\pm i\epsilon$, we have
\begin{align}
    q(-t\pm i\epsilon) =-\frac{t}{S-t}\pm \frac{i\epsilon S}{(S-t)^2}
    +\mathcal O(\epsilon^2)\,.
\end{align}
For $S>2b$, we have $|q(-t)|<1$. Thus, both hypergeometric
functions on the right-hand side admit convergent Gauss series. For $t\neq\tfrac{1}{2}$, these functions and their prefactors are analytic around $z=-t$. At $t=\tfrac{1}{2}$, the individual connection terms contain poles in parameters that cancel each other. So both limits $\mathcal G(-\tfrac{1}{2}\pm i0;S)$ remain finite. The discontinuity comes entirely from $q(z)^{3+2z}$ in the second term in the connection formula:
\begin{align}
    \lim_{\epsilon\to 0^+}q(-t\pm i\epsilon)^{3-2t\pm2i\epsilon} &=
    \left(\frac{t}{S-t}\right)^{3-2t} e^{\pm i\pi(3-2t)}\,.
\end{align}
Therefore, the discontinuity is given by
\begin{align}
    \operatorname{Disc}_t \mathcal G(-t;S) &=\frac{48i\Gamma(2t-3)\sin[\pi(3-2t)]}{\Gamma(1+t)^2} \left(\frac{t}{S-t}\right)^{3-2t} {}_2F_1\left(4-t,4-t;4-2t;-\frac{t}{S-t}\right)\,,\cr
    &=-\frac{48\pi i}{\Gamma(1+t)^2\Gamma(4-2t)}
    \left(\frac{t}{S-t}\right)^{3-2t}{}_2F_1\left(4-t,4-t;4-2t;-\frac{t}{S-t}\right)\,.
\end{align}
In the last step, we used $\Gamma(-x)\sin(\pi x)=-\pi/\Gamma(1+x)$. Including the gamma prefactor, we write
\begin{align}
    \mathcal N(-t\pm i0;S) &= -\frac{(2-t)(3-t)\Gamma(4-t)\Gamma(1+t)}
    {48t\,\Gamma(2-2t)}(S-t)^{-1-t}\,\mathcal G(-t\pm i0;S)\,,
\end{align}
giving the formula in \eqref{eq:disc_hypergeometric}.

\subsection{Saddle analysis of $\mathcal V_b(S,\tau)$}\label{app:LR_Vu_saddle}

We consider the contour integral
\begin{equation}
  \mathcal{V}_b(S,\tau) := \frac{1}{2\pi i}\left(\int_{-b-i\infty}^{-b-i0} + \int_{-b+i0}^{-b+i\infty}\right) dz\, \mathcal{I}(z;S,\tau)
  =: \mathcal{V}_{b,1}(S,\tau) + \mathcal{V}_{b,2}(S,\tau)\,,
\end{equation}
with $\frac{3}{4} < b < 1$, $S > 0$ large, and $\tau \in \mathbb{R} \setminus \{b\}$. For the saddle analysis, we first undo Pfaff's transformation\footnote{The Pfaff's transformation is
${}_2F_1(a,b;c;w)=(1-w)^{-a}
{}_2F_1(a,c-b;c;w/(w-1))$.} to express the hypergeometric function with argument $-S/z$ and then apply the connection formula at infinity to decompose
the integrand as $\mathcal I=\mathcal I_1+\mathcal I_2$ given below
\begin{align}
  \mathcal{I}_1(z;S,\tau) &= -\frac{\Gamma(-z)}{(z+\tau)\Gamma(1+z)}\, S^{z-1} H_1(z,S)\,, \\
  \mathcal{I}_2(z;S,\tau) &= \frac{(z+2)(z+3)\Gamma(-2z-3)\Gamma(z+4)}{2(z+\tau)\Gamma(1-z)\Gamma(2z+2)}\, S^{-4-z} z^{2z+2} H_2(z,S)\,,
\end{align}
and the Gauss hypergeometric functions are defined by
\begin{align}
  H_1(z,S) &:= {}_2F_1\!\left(1-z,\,-z-3;\,-2z-2;\,-\frac{z}{S}\right)\,, \\
  H_2(z,S) &:= {}_2F_1\!\left(z+4,\,z;\,2z+4;\,-\frac{z}{S}\right)\,.
\end{align}
For $\frac{3}{4} < b < 1$, the vertical line $\operatorname{Re}(z) = -b \in \left(-1, -\frac{3}{4}\right)$ lies strictly inside the pole-free strip between $z = -1$ and $z = -\frac{1}{2}$. Parameterizing $z = -b + it$ and using Schwarz reflection formula: $\mathcal{I}(\bar{z};S,\tau) = \overline{\mathcal{I}(z;S,\tau)}$, the upper and lower half-contours combine into
\begin{equation}
  \mathcal{V}_{b,k}(S,\tau) = \frac{1}{\pi} \operatorname{Re} \int_0^\infty dt\, \mathcal{I}_k(-b+it;S,\tau)\,, \qquad k \in \{1,2\}\,.
\end{equation}

Along $z = -b + it$, the factor $z^{2z+2}$ provides $e^{-\pi|t|}$ damping while $H_2(-b+it,S) \sim \left(\frac{1+\sqrt{1+z/S}}{2}\right)^{-2z}$ grows at most as $e^{+\frac{\pi}{2}|t|}$. Therefore, $|\mathcal{I}_2(-b+it;S,\tau)| \le C S^{b-4}(1+|t|)e^{-\frac{\pi}{2}|t|}$ for all $t$, so $\mathcal{V}_{b,2}(S,\tau) = \mathcal{O}(S^{b-4})$.
Thus, the leading large-$S$ contribution arises from $\mathcal{V}_{b,1}(S,\tau)$ in the stationary-phase regime $t \sim \sqrt{S} \gg 1$.

Using the Stirling's approximation
\begin{equation}
  \Gamma(\sigma \pm it)\Big|_{t\to \infty} = \sqrt{2\pi}\, |t|^{\sigma-1/2} e^{-\frac{\pi}{2}|t|}
  \exp\!\left[\pm i\left(t\ln t - t + \frac{\pi}{2}\left(\sigma - \frac{1}{2}\right)\right)\right] \left[1 + \mathcal{O}(t^{-1})\right]\,.
\end{equation}
we can write
\begin{equation}
  \frac{\Gamma(b-it)}{\Gamma(1-b+it)} = |t|^{2b-1} e^{-i(2t\ln t - 2t)}\left[1 + \frac{ib(1-b)}{t} + \mathcal{O}(t^{-2})\right]\,.
\end{equation}
Multiplying by $-\frac{1}{-b+\tau+it} = \frac{i}{t}\left[1 + \frac{i(\tau-b)}{t} + \mathcal{O}(t^{-2})\right] = t^{-1} e^{i\pi/2}\left[1 + \mathcal{O}(t^{-1})\right]$ and $S^{z-1} = S^{-b-1} e^{it\ln S}$, the prefactor in $\mathcal{I}_1$ takes the form
\begin{equation}\label{eq:prefactor_stirling}
  -\frac{\Gamma(-z)}{(z+\tau)\Gamma(1+z)}\, S^{z-1}
  = S^{-b-1} |t|^{2b-2} \exp\!\left[i\left(t\ln S - 2t\ln t + 2t + \frac{\pi}{2}\right)\right]\left[1 + \mathcal{O}(t^{-1})\right]\,.
\end{equation}

Expanding the Pochhammer ratio of the $j-$th term in the Gauss series representation of $H_1(z,S)$ at large $z$ yields
\begin{align}
  \frac{(1-z)_j(-z-3)_j}{(-2z-2)_j}\left(-\frac{z}{S}\right)^j
  = \left(\frac{z^2}{2S}\right)^j \left[1 + \mathcal{O}(z^{-1})\right]\,.
\end{align}
On $z=-b+it$, for sufficiently large $|t|$ and
$|z|\leq\kappa\sqrt S$, with $\kappa>0$ fixed we obtain
\begin{align}\label{eq:H1_gaussian}
  H_1(-b+it, S) = e^{-\frac{t^2}{2S}}\left[1 + \mathcal{O}(|t|^{-1})\right]\,.
\end{align}

Combining \eqref{eq:prefactor_stirling} and \eqref{eq:H1_gaussian}, the integral for $\mathcal{V}_{b,1}(S,\tau)$ in the region $t \sim \sqrt{S}$ is governed by
\begin{align}
  \mathcal{V}_{b,1}(S,\tau) \sim \frac{S^{-b-1}}{\pi} \operatorname{Re} \int_0^\infty dt\, t^{2b-2} e^{-\frac{t^2}{2S}} e^{i\Psi(t)}\left[1 + \mathcal{O}(S^{-1/2})\right]\,,
\end{align}
where the phase $\Psi(t)$ and its first two derivatives are
\begin{align}
  \Psi(t) = t\ln S - 2t\ln t + 2t + \frac{\pi}{2}\,, \qquad
  \Psi'(t) = \ln S - 2\ln t\,, \qquad
  \Psi''(t) = -\frac{2}{t}\,.
\end{align}
The unique stationary point in $(0,\infty)$ lies at
\begin{align}
  \Psi'(t_*) = 0 \implies t_* = \sqrt{S}\,,
\end{align}
at which the logarithmic terms in $\Psi(t_*)$ cancel identically:
\begin{align}
  \Psi(t_*) = 2\sqrt{S} + \frac{\pi}{2}\,, \qquad
  \Psi''(t_*) = -\frac{2}{\sqrt{S}} < 0\,, \qquad
  e^{-\frac{t_*^2}{2S}} = e^{-1/2}\,.
\end{align}
Therefore, the standard stationary phase formula yields
\begin{align}
  \mathcal{V}_{b,1}^{\mathrm{saddle}}(S,\tau)
  &= \frac{S^{-b-1}}{\pi} \left(\sqrt{S}\right)^{2b-2} e^{-1/2} \sqrt{\frac{2\pi}{|\Psi''(t_*)|}}\,
     \operatorname{Re}\exp\left[i\left(\Psi(t_*) - \frac{\pi}{4}\right)\right] \nonumber \\
  &= \frac{2}{\sqrt{e\pi}}\, S^{-7/4} \cos\left(2\sqrt{S} + \frac{\pi}{4}\right)\,,
\end{align}
and thus establishing \eqref{eq:LR_oscillatory_remainder} for $\frac{3}{4} < b < 1$.

\section{Kernel relation}\label{app:kernel_relations}

The leading Regge (LR) and subleading Regge (SLR) contributions select the very first two family of intermediate states from the factorization sum of the full string amplitude. Their expressions in terms of Jacobi and Legendre polynomials nevertheless suggest a relation at the kernel level exists. We make this relation explicit by constructing a differential operator that maps the LR polynomial at mass level $n$ to the SLR polynomial at level $n+1$. 
Together with the corresponding changes in the pole position, kinematic argument, and normalization, this yields a map between these kernels. In particular, we seek an operator satisfying
\begin{align}
    \mathcal{O}_n P_{n-1}^{(0,4)}(x)=P_n(x)\,,
\end{align}
where $n = \alpha'm_n^2$ denote the mass level. To determine $ \mathcal{O}_n$,
note that the Jacobi polynomial satisfies following recurrence relations
\begin{align}
    \mathcal B_{p,\beta}P_p^{(0,\beta)}(x) &=P_p^{(0,\beta-1)}(x)\,, \\
    \mathcal V_{p,\beta}P_p^{(0,\beta)}(x) &=P_{p+1}^{(0,\beta-1)}(x)\,,
\end{align}
where 
\begin{align}
    \mathcal B_{p,\beta} &:=\frac{(1-x^2)\partial_x+p(1+x)+2\beta}{2(p+\beta)}\,,\\
    \mathcal V_{p,\beta} &:=\frac{(x^2-1)\partial_x+(p+\beta+1)x+p-\beta+1}{2(p+1)}\,.
\end{align}
Thus $\mathcal V$ raises the polynomial degree while lowering the second Jacobi parameter, whereas $\mathcal B$ lowers that parameter at fixed degree:
\begin{align}
    P_{n-1}^{(0,4)} \xrightarrow{\;\mathcal V_{n-1,4}\;}
    P_n^{(0,3)} \xrightarrow{\;\mathcal B_{n,3}\;}
    P_n^{(0,2)} \xrightarrow{\;\mathcal B_{n,2}\;}
    P_n^{(0,1)} \xrightarrow{\;\mathcal B_{n,1}\;} P_n \,.
\end{align}
Therefore
\begin{align}
    \mathcal O_n :=\mathcal B_{n,1}\mathcal B_{n,2} \mathcal B_{n,3}\mathcal V_{n-1,4}\,,
    \qquad \mathcal O_n P_{n-1}^{(0,4)} = P_n \,.
\end{align}
After stripping the common factor $\alpha'^2[12]^2\langle34\rangle^2$ from the level-sum expressions, we define the dimensionless kernels
\begin{align}
    \widehat{\mathcal K}_n^{\mathrm{LR}}(x_1) :=a_nP_{n-1}^{(0,4)}(x_1)\,,\qquad
    \widehat{\mathcal K}_{n+1}^{\mathrm{SLR}}(x_2):=b_{n+1}P_n(x_2)\,,
\end{align}
where subscripts denote intermediate massive levels. The polynomial arguments depend on the exchanged mass and the channel kinematics:
\begin{align}
    x_1:=\frac{2s}{\alpha'm_{n,\mathrm{LR}}^2}-1
        =\frac{2s}{n}-1,
    \qquad
    x_2:=\frac{2\tau}{\alpha'm_{n,\mathrm{SLR}}^2}-1
        =\frac{2\tau}{n+1}-1,
\end{align}
where $m_{n,\mathrm{SLR}}^2=(n+1)/\alpha'$ denotes the mass of the SLR state with spin $n$. The coefficients can be read from \eqref{eq:leading_regge_jacobi_kernel} and \eqref{eq:subleading_regge_legendre_kernel}
\begin{align}
    a_n=\frac{n^{n-2}(n+2)(n+3)}{2\Gamma(2n+2)}\,, \qquad b_{n+1}=-144\, \frac{(-4)^n(n+1)^{n-1}}{\Gamma(2n+1)}\,.
\end{align}
Evaluating both kernels at a common independent variable $x$, the differential relation relating the two kernels becomes
\begin{align}
    \widehat{\mathcal K}_{n+1}^{\mathrm{SLR}}(x)  =\kappa_n\mathcal O_n
    \widehat{\mathcal K}_n^{\mathrm{LR}}(x)\,,
\end{align}
with
\begin{align}
    \kappa_n:=\frac{b_{n+1}}{a_n} =-288 \frac{(-4)^n(2n+1)(n+1)^{n-1}}{n^{n-2}(n+2)(n+3)}\,.
\end{align}
The channel-dependent kernels are recovered by evaluating the resulting functions at $x=x_1$ and $x=x_2$, respectively.

\end{appendix}

\bibliographystyle{JHEP}
\bibliography{string}

\end{document}